\documentclass[prx,twocolumn, showpacs, floatfix,tightenlines amsmath, amssymb, 
preprintnumbers, superscriptaddress, longbibliography]{revtex4-1}

\usepackage{graphicx}
\usepackage{color}
\usepackage{dcolumn}
\usepackage{bm}
\usepackage{hyperref}
\usepackage[nolist,nohyperlinks]{acronym}
\usepackage{CJK}
\usepackage{braket}
\usepackage{nicefrac}
\usepackage{amsmath}

\definecolor{cblue}{RGB}{19,107,192}
\hypersetup{colorlinks=true,linkcolor=cblue,citecolor=cblue,urlcolor=cblue}

\newcommand{\FeIon}{Fe$^{2+}$}
\newcommand{\FSS}{FeSc$_2$S$_4$}
\newcommand{\Fdm}{$Fd\bar{3}m$}
\newcommand{\Ibm}{$I\bar{4}m2$}

\newcommand{\mub}{$\mu_B$}
\newcommand{\wn}{cm$^{-1}$}

\newcommand{\bfpsi}{\mbox{\boldmath$\psi$}}

\begin{document}
\title{Antiferromagnetic and Orbital Ordering on a Diamond Lattice Near Quantum 
    Criticality}
\author{K.W. Plumb}
\affiliation{Institute for Quantum Matter and Department of Physics and 
    Astronomy, The Johns Hopkins University, Baltimore, MD 21218, USA}
\author{Jennifer Morey}
\affiliation{Institute for Quantum Matter and Department of Physics and 
    Astronomy, The Johns Hopkins University, Baltimore, MD 21218, USA}
\affiliation{Department of Chemistry, The Johns Hopkins University, Baltimore, 
    MD 21218, USA}
\author{J. A. Rodriguez-Rivera}
\affiliation{NIST Center for Neutron Research, National Institute of Standards 
    and Technology, Gaithersburg, MD 20899, USA}
\affiliation{Department of Materials Science and Engineering, University of 
    Maryland, College Park, MD 20742, USA}                                             
\author{Hui Wu}
\affiliation{NIST Center for Neutron Research, National Institute of Standards 
    and Technology, Gaithersburg, MD 20899, USA}
\author{A. A. Podlesnyak}
\affiliation{Quantum Condensed Matter Division, Oak Ridge National Laboratory, 
    Oak Ridge, Tennessee 37831-6473, USA}
\author{T. M. McQueen}
\affiliation{Institute for Quantum Matter and Department of Physics and 
    Astronomy, The Johns Hopkins University, Baltimore, MD 21218, USA}
\affiliation{Department of Chemistry, The Johns Hopkins University, Baltimore, 
    MD 21218, USA}
\affiliation{Department of Materials Science and Engineering, The Johns Hopkins 
    University, Baltimore, MD 21218, USA}
\author{C. L. Broholm}
\affiliation{Institute for Quantum Matter and Department of Physics and 
    Astronomy, The Johns Hopkins University, Baltimore, MD 21218, USA}
\affiliation{NIST Center for Neutron Research, National Institute of Standards 
    and Technology, Gaithersburg, MD 20899, USA}		
\affiliation{Quantum Condensed Matter Division, Oak Ridge National Laboratory, 
    Oak Ridge, Tennessee 37831-6473, USA}				
\date{\today}

\begin{abstract}
    
We present neutron scattering measurements on powder samples of the spinel 
\FSS{} that reveal a previously unobserved magnetic ordering transition 
occurring at 11.8(2)~K. Magnetic ordering occurs subsequent to a subtle 
cubic-to-tetragonal structural transition which  distorts Fe coordinating 
sulfur tetrahedra lifting the orbital degeneracy. The application of 1~GPa 
hydrostatic pressure appears to destabilize this N\'eel state, reducing the 
transition temperature to 8.6(8)~K and redistributing magnetic spectral weight 
to higher energies. The relative magnitudes  of ordered $\langle m 
\rangle^2\!=\!3.1(2)$ and fluctuating moments $\langle \delta m 
\rangle^2\!=\!13(1)$ show that the magnetically ordered ground state of \FSS{} 
is drastically renormalized and in proximity to criticality.
\end{abstract}

\maketitle

In a quantum spin liquid quantum fluctuations compete with, and ultimately 
overwhelm, any tendency towards the formation of a classical long range 
magnetic order, effectively melting the staggered magnetization at zero 
temperature.  The search for this precarious state of matter in a real material 
is now a decades old preeminent theme of condensed matter physics 
\cite{PALee:2008}. In insulating antiferromagnets, research has largely been 
concentrated on geometrically frustrated materials, where the underlying 
lattice structure results in a  competition amongst magnetic exchange 
interactions. The result is an extensive degeneracy that promotes quantum 
fluctuations and precludes the development of a staggered magnetization 
\cite{Balents:10}. But frustration is not the only game in town.  Other degrees 
of freedom may enhance magnetic fluctuations.  For example,  orbital degeneracy 
which can in turn enhance quantum fluctuations and suppress magnetic order 
\cite{Feiner:1997}. The orbital degrees of freedom may even
remain disordered in the presence of a staggered magnetization forming an 
orbital liquid state \cite{Kitaoka:1998,Keimer:2000,Khaliullin:2000},
or both spins and orbitals may fail to develop temperature independent 
correlations in a so-called spin-orbital liquid state \cite{Nakatsuji:2012, 
    Corboz:2012}. 

Spinel compounds, AB$_2$X$_4$, with magnetic ions occupying the A-site diamond 
sublattice form a simple, and therefore important, three dimensional frustrated 
lattice where competing exchange interactions promote unconventional magnetic 
ground states stabilized by fluctuations \cite{Krimmel:2006, Bergman:2007, 
    Bernier:2008, Krimmel:2009, MacDougall:2011}.  Amongst the A-site spinels, 
\FSS{} holds a special place as the Fe$^{2+}$ (3d$^6$) ions are both 
magnetically and orbitally active.  The tetrahedral crystal field of the A-site 
environment splits the Fe 3d manifold into a lower $e$ doublet and higher 
energy $t_{2}$ triplet, while Hund's coupling yields a high spin (S=2) 
configuration, with a single hole occupying the $e$ doublet.  Thus, the 
\FeIon{} ion in \FSS{} is Jahn-Teller active yet, surprisingly, there have been 
no reported observations of any structural distortion or magnetic ordering down 
to 50~mK. In particular, while the magnetic susceptibility exhibits Curie-Weiss 
like behaviour with $\theta_{CW}\!=\!-45$~K \cite{Fritsch:2004,Buttgen:2004}; 
there is no clear indication of a phase transition. Moreover the specific heat 
only exhibits a broad peak with maximum centered at 8~K.  Neutron scattering 
measurements on polycrystalline samples conducted so far have not provided
evidence for magnetic ordering. The data show intense dispersive spin 
excitations emerging from  momentum transfer corresponding to the cubic (100) 
position with an apparent excitation gap of $\sim$0.2 meV \cite{Krimmel:2005}.  

Based on these experimental observations it was proposed that \FSS{} is a 
spin-orbital liquid: an exotic state arising from competition between on site 
spin-orbit coupling and Kugel-Khomskii exchange in the Fe diamond sub-lattice 
\cite{Chen:2009, Chen:2009b}. While spin and orbital ordering is favoured by 
the exchange interactions, atomic spin orbit coupling favours local 
spin-orbital singlets and \FSS{} is predicted to lie near the quantum critical 
point separating the spin-orbital singlet phase from a magnetically and 
orbitally ordered state.  This hypothesis appears supported by terahertz and 
far infrared optical spectroscopy measurements which  have revealed low energy 
(4.5~meV) excitations consistent with a so-called spin-orbiton 
\cite{Mittelstadt:2015, Laurita:2015, Ish:2015}, an excitation of entangled 
spin and orbital degrees of freedom. In this scenario, varying the magnetic 
exchange interactions relative to the spin orbit coupling through the 
application of hydrostatic pressure might drive \FSS{} through the quantum 
critical point, stabilizing antiferromagnetism \cite{Chen:2009}.
\begin{figure}[t!]
    \includegraphics[]{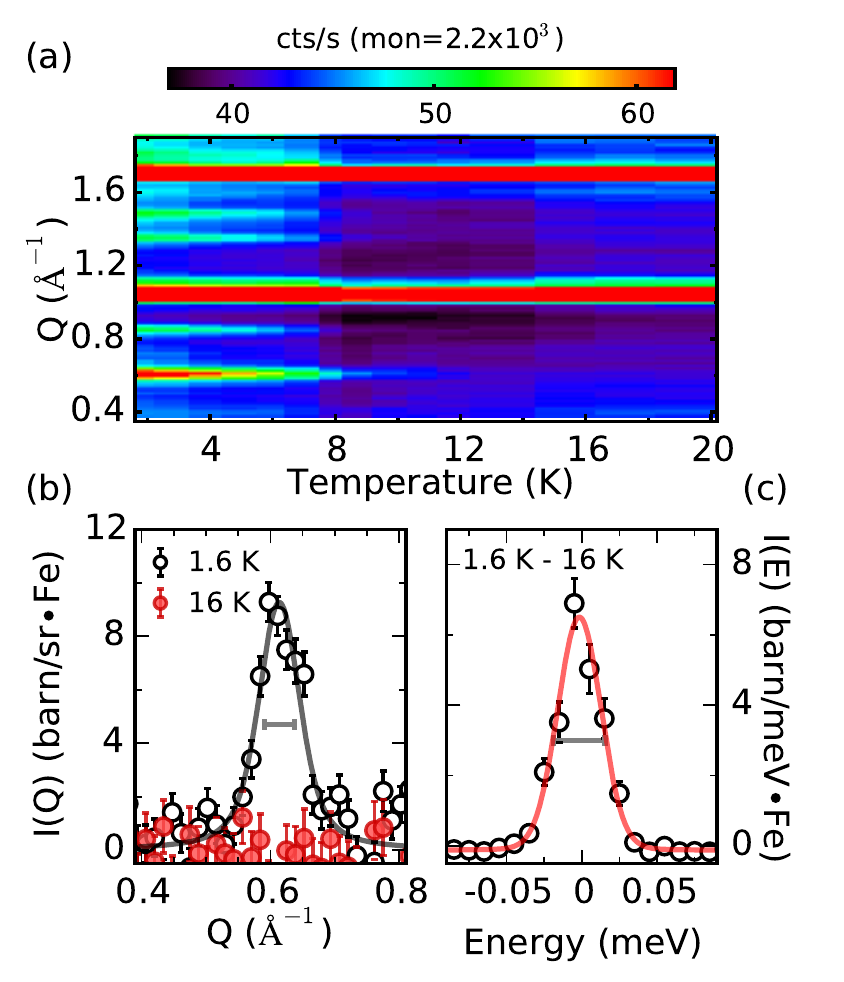}
    \caption{\label{fig:TDepBragg} (a) Overview of the temperature dependent  
        neutron diffraction intensity in \FSS{} as measured on MACS. (b) 
        Magnetic Bragg reflection measured on CNCS integrated between 
        $-0.018\!<\!E\!<\!0.018$~meV, an identical constant background has been 
        subtracted from low and high temperature data sets. (c) Energy 
        dependence of magnetic Bragg reflection integrated over 
        $0.55\!<\!Q\!<0.65$ \AA$^{-1}$. Horizontal bars in (b) and (c) indicate 
        the instrumental resolution as determined from corresponding cuts 
        through the nuclear (111)  Bragg peak at 1.04~\AA$^{-1}$. Error bars 
        represent one standard deviation.}
\end{figure}

Here we show that contrary to the spin-orbital liquid scenario, \FSS{} 
undergoes a magnetic ordering transition concomitant with the maximum observed 
in specific heat. Our data also indicate a subtle tetragonal, Jahn-Teller-like, 
distortion occurring at higher temperatures which lifts the orbital degeneracy 
so that the low temperature dynamics can be described in terms of a spin-only 
Hamiltonian.  Additionally, we find that the application of hydrostatic 
pressure of 1~GPa reduces the magnetic ordering temperature and shifts 
inelastic spectral weight to higher energies, apparently destabilizing the 
ordered state.  The N\'eel state in \FSS{} is strongly renormalized with a
sublattice magnetization that is reduced to 44(1)\% of the full saturation 
moment and 40(2)\% of the inelastic spectral weight associated with non-spin 
wave fluctuations. These results show that \FSS{} is indeed near a quantum 
critical point, but surprisingly, on the magnetically ordered side of this 
critical point.

\section*{Methods}
Powder samples of \FSS{} were synthesized by solid state reaction of elemental 
Fe, Sc and S  in an evacuated quartz tube. All materials were handled in a 
glove box using standard air free techniques. The sample was initially 
characterized by powder x-ray diffraction at 300~K using a Cu-$K\alpha$ x-ray 
diffractometer.  Additional high resolution synchrotron x-ray diffraction 
measurements were acquired $T=$100~K using the powder diffractometer 11-BM 
located at the Advanced Photon Source, these are described in 
Appendix~\ref{sec:S_XRD}.   

Specific heat measurements were carried out on a cold pressed pellet of \FSS{} 
using a Quantum Designs PPMS. The MACS spectrometer at the NIST center for 
Neutron Research was used for neutron scattering measurements.  For 
measurements at ambient pressures, the 0.8~g sample of \FSS{} was mounted in a 
standard Al sample can.  Measurements under hydrostatic pressure were conducted 
with the sample loaded into a stainless steel pressure cell and pressurized to 
1~GPa with He gas. The lattice parameter of a pyrolytic graphite standard 
loaded into the pressure cell just above the sample was used to monitor the 
quasi-hydrostatic pressure at low temperature.  MACS was operated with a fixed 
final neutron energy of either 3.7~meV or 2.4~meV with appropriate combinations 
of Be and BeO filters located before and after the sample to suppress higher 
order neutron contamination. No incident beam filter was used  for measurements 
with incident energies above 5~meV, and data for these energies were corrected 
to account for contributions to the monitor count rate from higher order 
neutrons \cite{Stock:04}.  To compare absolute intensities between the 0~kbar 
and 10~kbar measurements data collected in the neutron attenuating pressure 
cell were corrected for the measured energy dependent neutron transmission of 
the cell.  Neutron diffraction measurements on the same sample were performed 
on BT-1 at NIST utilizing a Cu(311) monochrometer, neutron wavelength 1.54~\AA,  
and 60' collimation before the monochrometer.  A set of high resolution neutron 
scattering measurements were conducted on the CNCS spectrometer located at the 
Spallation Neutron Source at Oak Ridge National Lab \cite{Ehlers:2011}.  For 
these measurements a separate 0.45~g powder sample was prepared and CNCS was 
operated with a fixed incident neutron energy of 1.55~meV and a resulting 
elastic energy resolution of 0.034~meV (FWHM).  For all scattering measurements 
background signal contributions from the sample environment were subtracted and 
signal count rates were converted to absolute values of the scattering 
cross-section using the integrated intensity from a nuclear Bragg reflection 
\cite{Lee:1997}.  

We explicitly define the normalized magnetic neutron scattering intensity as 
used in this work to aid accurate discussion of sum-rules presented later.  The  
magnetic neutron scattering cross section for a powder sample as a function of 
momentum transfer $Q$, and energy transfer $E$ is expressed as
\begin{align}
    \frac{d^2\sigma}{d\Omega dE^{\prime}} = \frac{N}{\hbar} \frac{k_f}{k_i} 
    r_0^2 e^{-2W(Q)} 2\mathcal{\tilde{S}}\left(Q,E\right),
\label{eq:Mag_CrossSect}
\end{align}
where N is the number of unit cells, $k_i$ and $k_f$ are the incident and final 
neutron wavevectors respectively, $r_0\!=\!0.539\times10^{-12}$~cm is the 
magnetic scattering length, and $e^{-2W(Q)}$ is the Debye Waller factor  which 
we set to unity for low T and low Q. The spherically averaged dynamic structure 
factor is given by
\begin{align}
    \mathcal{\tilde S}(Q,E)  &= 
    \left|\frac{g}{2}f\left({Q}\right)\right|^2\times \nonumber \\
    &  \int{\frac{d\Omega_{Q}}{4\pi}
\sum_{\alpha \beta} \left( \delta_{\alpha \beta}\!-\!  
    \hat{Q}_{\alpha}\hat{Q}_{\beta}\right)\mathcal{S}^{\alpha\beta}\left({\bf 
        Q},E\right)},
\label{eq:Mag_CrossSect}
\end{align}
where  $g$ is the Land\'e g-factor and  $f\left(Q\right)$ is the magnetic form 
factor. In this work we report the normalized inelastic neutron scattering 
intensity per Fe
\begin{equation}
\tilde{I}\left(Q,E\right)=\frac{k_i}{k_f} \frac{d^2\sigma}{d\Omega 
    dE^{\prime}}.
\label{eq:IQE}
\end{equation}

Rietveld refinement of neutron and synchrotron diffraction data was carried out 
using the General Structure Analysis system (GSAS) complimented with the 
SARA{\it h} program~\cite{Wills:Sarah}.

\section*{Results}
Our main experimental finding is presented in Fig.~\ref{fig:TDepBragg} (a) 
which displays a false color map of the temperature dependent neutron 
diffraction intensity as measured on MACS. The most prominent features are two 
structural Bragg peaks at $Q\!=\!1.03$ \AA$^{-1}$ and $1.69$ \AA$^{-1}$  that 
saturate the color scale and are independent of temperature below 20~K.  More 
importantly, there are five temperature dependent Bragg reflections which 
emerge in unison below 12~K. Located at: 0.6~\AA$^{-1}$, 0.84~\AA$^{-1}$, 
1.33~\AA$^{-1}$, 1.46~\AA$^{-1}$, and 1.77 \AA$^{-1}$, these peaks were not 
reported for \FSS{}, although published diffraction data are not inconsistent 
with these findings as their sensitivity is insufficient to access this weak 
signal. In the following we will argue that these Bragg peaks arise from a 
magnetic ordering transition which is enabled by a higher temperature orbital 
ordering transition. Details of the lowest Q magnetic Bragg reflection are 
shown in Fig.~\ref{fig:TDepBragg} (b) and (c). We fit the high resolution data 
using a Lorentzian profile convolved with a Gaussian instrumental resolution of 
0.045 \AA$^{-1}$ full width at half maximum (FWHM). This reveals an intrinsic 
Lorentzian half width at half maximum of $\kappa\!=\!0.019(6)$~\AA$^{-1}$ and 
corresponding correlation length of 
$\xi\!=\!\nicefrac{1}{\kappa}\!=\!53(16)$~\AA.  Similarly the energy dependence 
of the Q-integrated peak was fit to a quasielastic Lorentzian profile convolved 
with the Gaussian instrumental resolution to yield a lower bound on the 
correlation time of $\tau\!>\!0.4$~ns. The finite magnetic correlation length 
might result from exchange disorder built into the system at the higher 
temperature orbital occupational ordering transition as discussed below. 

\subsection*{Specific Heat}
\begin{figure}[t!]
    \includegraphics[]{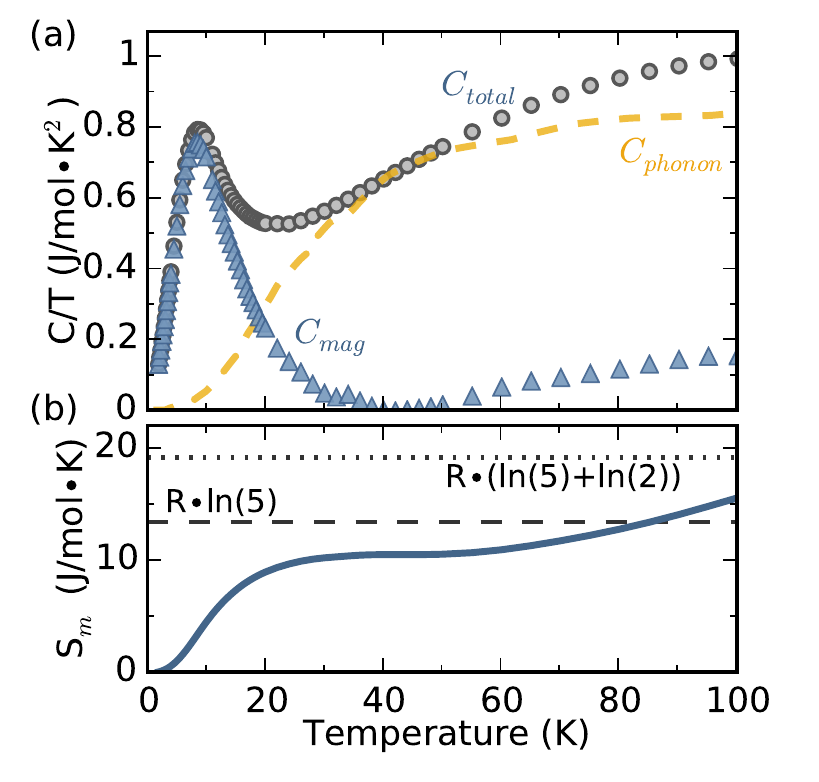}
\caption{\label{fig:SpecificHeat}
    Specific heat for powder samples of \FSS{} used in this work.  (a) Total 
    specific heat (grey circles) and estimated magnetic contribution (blue 
    triangles). Dashed yellow line is the estimated phonon contribution.  (b) 
    Estimated magnetic entropy of \FSS{}.}
\end{figure}

The specific heat of our powder sample shown in Fig~\ref{fig:SpecificHeat} has 
qualitative similarities to previous reports, but there are many quantitative 
differences \cite{Fritsch:2004,Buttgen:2004}.  No sharp anomaly signifying a 
phase transition is visible around 12~K and instead there is a very broad 
maximum in C/T at 8~K. This slightly higher than the 6~K maximum reported in 
earlier studies \cite{Fritsch:2004,Buttgen:2004}.  The turnover at 8~K is also 
considerably shaper in temperature; both of these observations are consistent 
with a reduced disorder in the samples studied here. Phonon contributions to 
the specific heat were estimated by appropriately scaling the measured specific 
heat of the non-magnetic isostructural compound CdIn$_2$S$_4$ as described in 
Appendix~\ref{sec:phonon_cp}. The phonon contribution, shown as a dashed line 
in Fig.~\ref{fig:SpecificHeat} (a), was subtracted from the total specific heat 
to yield the magnetic component.  From this we calculate the magnetic entropy, 
$S_m$ shown in Fig.~\ref{fig:SpecificHeat} (b).  Rather than exceeding the 
$R\ln{(5)}$ entropy available for a $S\!=\!2$ system, the magnetic entropy 
plateaus at approximately 70~\% of the available spin entropy, approaching 
$R\ln{(5)}$ at 80~K.  We find no evidence for orbital fluctuations in the 
magnetic entropy at low temperatures, only above 80~K does the entropy begin to 
recover the additional $R\ln{(2)}$ available for a two-fold orbital degeneracy 
and our analysis indicates that the orbital degeneracy in \FSS{} is entirely 
quenched below $\sim$80~K.  

\subsection*{Orbital Degeneracy Breaking}
\begin{figure}[t!]
    \includegraphics[]{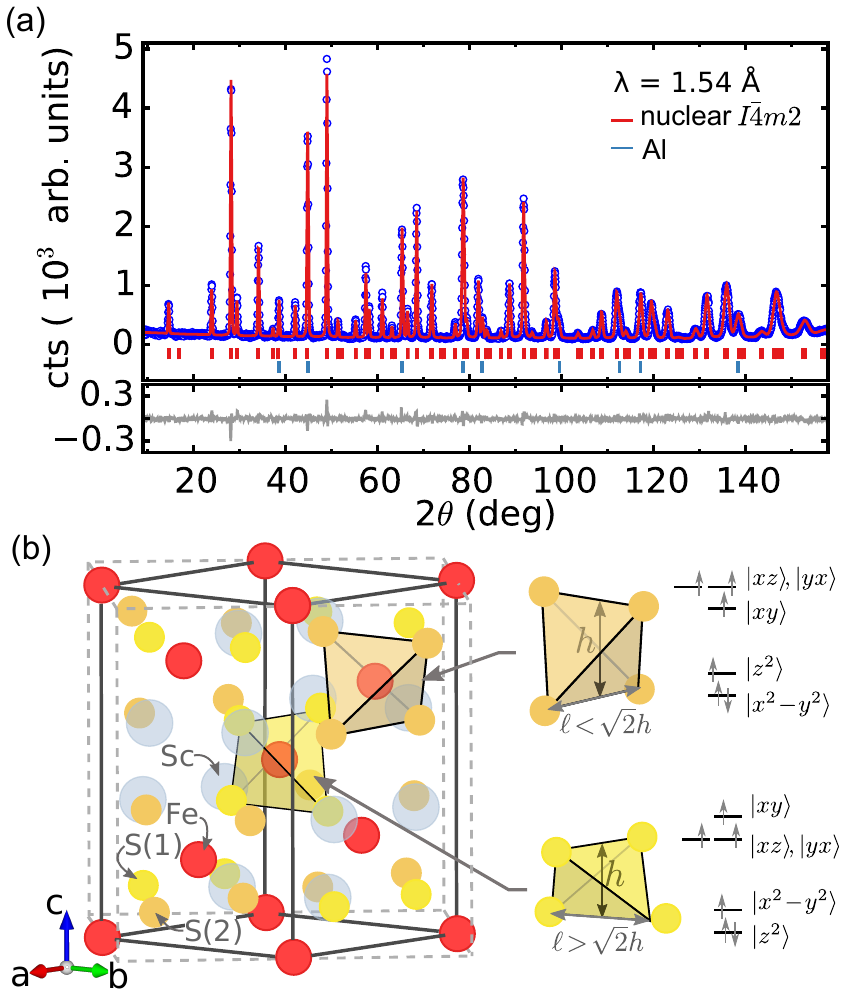}
\caption{\label{fig:Structure} (a) Neutron powder diffraction pattern for 
    \FSS{} measured on BT-1 at 20~K, open symbols are observed intensities and 
    solid red line is a Rietveld refinement of the nuclear structure to the 
    $I\bar{4}m2$ space group. (b) Proposed $I\bar{4}m2$ crystal structure of 
    \FSS{}, the refined lattice parameters are $a\!=\!7.434(1)$~\AA{} and 
    $c\!=\!  10.493(1)$~\AA{}.  There are two crystallographically distinct 
    sulfur sites, labelled S(1) and S(2), resulting in two different Fe$^{2+}$ 
    coordinating tetrahedra alternately expanded and contracted along the 
    tetragonal (001) direction.  Fe$^{2+}$ orbital occupancies for small 
    distortions of the S tetrahedra are shown on the right.  The parent cubic 
    cell is shown as a light grey dashed outline.}
\end{figure}

An orbitally non-degenerate state at low temperatures is at odds with the 
previously accepted crystal structure for \FSS{} we shall thus first re-examine 
the low temperature crystal structure.  The T=20~K neutron diffraction pattern 
of \FSS{} is shown in Fig.~\ref{fig:Structure} (a).  All peaks can be indexed 
by the cubic \Fdm{} space group and no peak splitting characteristic of a
symmetry lowering structural distortion is apparent.  Nonetheless, Rietveld 
refinement of our diffraction data using the \Fdm{} space group is not 
satisfactory as it yields a systematic discrepancy between measured and 
calculated peak intensities. Amongst the maximal subgroups of \Fdm{} we find a 
significantly improved description of the data using the tetragonal \Ibm{} 
space group. Specifically, and as described more fully in Appendix 
\ref{sec:struct_models}, the goodness of fit $\chi^2$ is reduced from $1.87$ to 
$1.30$ in going from \Fdm{} to \Ibm{}.  Fig.~\ref{fig:Structure} (b) shows the 
new unit cell and the resulting Rietveld refinement is shown 
in~Fig.~\ref{fig:Structure} (a),  details of the refinement are presented in 
Table~\ref{tab:tetrag_struct}.

Within our measurement resolution the \FSS{} unit cell maintains its cubic 
metric and the tetragonal distortion is limited to 
$\nicefrac{c}{a_{c}}\!=\!0.998(2)$, where the parent cubic cell lattice 
parameter $a_{c}\!=\!\sqrt{2}a_{t}$ and $a_{t}$ is the tetragonal cell lattice 
parameter.  The \Ibm{} unit cell contains \emph{two} crystallographically  
distinct Fe sites, each site comprising a  \emph{centered tetragonal} (CT) 
sublattice.  Importantly, the Fe-coordinating sulfur tetrahedra for each 
respective sublattice are alternately compressed or elongated along the (001) 
direction, as shown in Fig.~\ref{fig:Structure} (b).  The tetragonal distortion 
thus relaxes the tetrahedral symmetry around each iron site, lifting the 
orbital degeneracy of the Fe $e$ orbitals and results in a hole occupying the
$\ket{z^2}$ orbital for one sublattice and the $\ket{x^2-y^2}$ orbitals for the 
other. From an energetic perspective, this particular distortion minimizes 
macroscopic strain while fully quenching the orbital moment. High resolution 
x-ray diffraction measurements show that the average structure is cubic at 
least to 100~K, but reveal a large anisotropic microstrain broadening of the 
diffraction peaks possibly indicating an incipient structural transition (See 
Appendix~\ref{sec:S_XRD}. Further neutron and x-ray diffraction experiments 
that span the relevant temperature regime where the phase transition may occur 
are under way. 

While no deviations from cubic symmetry have been reported thus far, the 
particular low temperature structure we report resolves a number of 
inconsistencies in the literature. First, M\"ossbauer spectroscopy consistently 
indicates two distinct Fe sites \cite{Brossard:1976,Son:2008}. This has been 
ascribed to result from Fe-Sc site mixing at the level of 30~\%\cite{Son:2008}, 
but seems an extremely unlikely scenario given the disparate ionic radii and 
charge of Fe$^{2+}$ and Sc$^{3+}$, and is furthermore inconsistent with neutron 
diffraction, which offers excellent contrast between Fe and Sc.  Even at room 
temperature the M\"ossbauer spectra necessitate a second Fe site in small 
fractions and below $\sim$50~K the quadrupolar splitting from each respective 
site diverges. These observations seem consistent with a thermally fluctuating, 
or dynamic, distortion which dramatically slows down below 50~K.  Second, 
optical measurements in the far infrared show signatures of symmetry breaking 
\cite{Reil:2002, Mittelstadt:2015}.  Two bands around 467~\wn{} are observed in 
the 300~K absorption spectra which are forbidden in the \Fdm{} space group and 
have lacked a satisfactory explanation \cite{Reil:2002}.  The extra absorption 
bands indicate a local distortion is present, at least dynamically, even at 
room temperature.
\begin{table}[t!]
\caption{Atomic parameters for the proposed low temperature tetragonal 
    structure of \FSS{} at 20~K.  The space group is $I\bar{4}m2$ (119) with 
    lattice parameters $a\!=\!7.434(1)$~\AA{} and $c\!=\!  10.493(1)$~\AA{}.  
    Reitveld refinements resulted in a $\chi^2$ of 1.33 and 
    $R_{Bragg}\!=\!5.25$\%.}
\begin{tabular}{lcccccc}
\hline
\hline
Atom & Wyck. site & x & y & z & Occ &B$_{iso}$\\
\hline
Fe(1) & 2a & 0.0000   & 0.0000 & 0.0000 & 1.0  & 0.183(4) \\
Fe(2) & 2c & 0.0000   & 0.5000 & 0.2500 & 1.0  & 0.183(4) \\
S(1) & 8i  & 0.2618(3)   & 0.0000 & 0.1283(8) & 1.0  & 0.216(4) \\
S(2) & 8i  & 0.2434(8)  & 0.0000 & 0.6207(8) & 1.0  & 0.216(4) \\
Sc & 8i  & 0.7541(3)   & 0.0000 & 0.3741(6) & 1.0 & 0.188(4) \\
\hline
\hline
\end{tabular}
\label{tab:tetrag_struct}
\end{table}

Any signatures of a structural distortion in our diffraction data are subtle as 
the low temperature structure remains metrically cubic.  A small distortion is, 
however, generally consistent with expectations for tetrahedrally coordinated 
Fe$^{2+}$.  Indeed, any structural anomaly associated with orbital ordering in 
the related compound FeCr$_2$S$_4$ was only apparent through careful line shape 
analysis of high resolution x-ray synchrotron diffraction data 
\cite{Tsurkan:2010}.  Furthermore, we shall see that for \FSS{} the presence of 
a tetragonal distortion proves essential to consistently describe both the 
observed magnetic ordering and excitation spectra.

\subsection*{Magnetic Ordering}
\begin{figure}[t!]
    \includegraphics[]{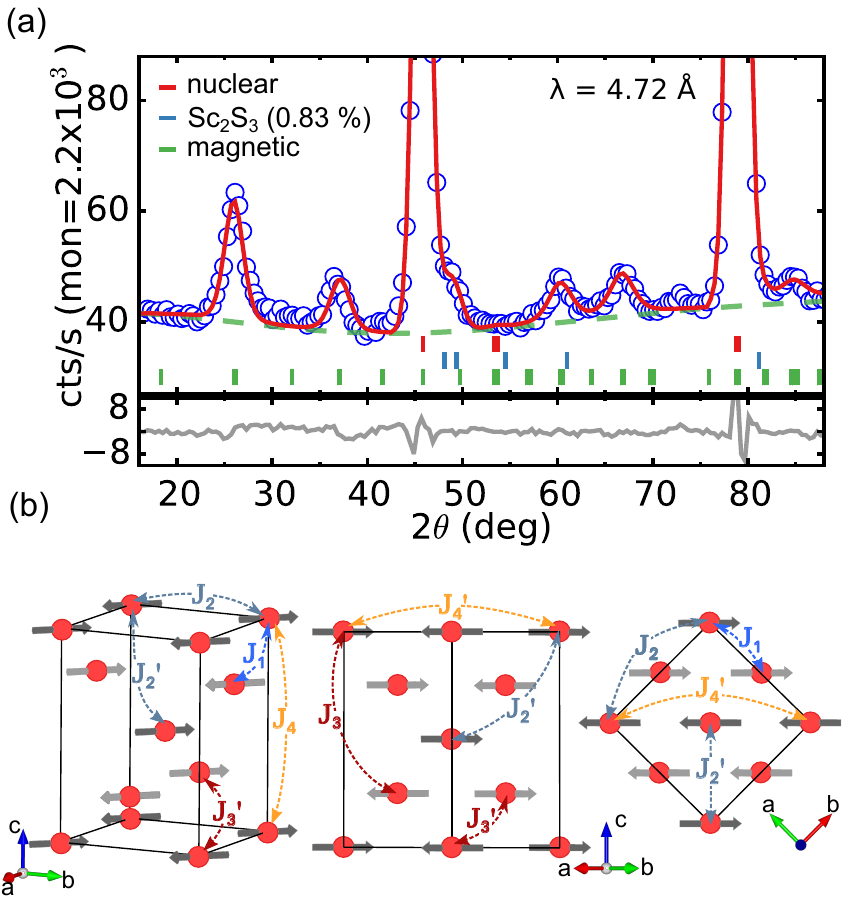}
\caption{\label{fig:MagStructure}  (a) Neutron  powder diffraction collected on 
    MACS at 1.6~K.  Open symbols are observed intensities, solid line is the 
    calculated intensity for the magnetic structure shown in (b), and dashed 
    green line is background.  (b) Proposed magnetic structure of \FSS{} shown 
    in perspective view, projection along the (110), and (001) directions.  
    Identical moments on crystallographically distinct Fe sites are drawn in 
    light and dark grey to highlight the two CT sublattices.}
\end{figure}

The magnetic peaks may be indexed in the tetragonal unit cell using a single 
propagation vector of either $q_m\!=\!(0,0,1)$ or 
$q_m\!=\!(\nicefrac{1}{2},\nicefrac{1}{2},0)$.  The resulting magnetic 
structure and refinement are shown in figures~\ref{fig:MagStructure} (a) and 
(b).  Refinements for magnetic models using a propagation vector of either 
$(0,0,1)$ or $(\nicefrac{1}{2},\nicefrac{1}{2},0)$ are indistinguishable so 
that a unique solution cannot be determined from diffraction alone. However, as 
will be discussed below, the powder averaged inelastic magnetic neutron 
scattering cross section can only be accounted for based on the 
$(\nicefrac{1}{2},\nicefrac{1}{2},0)$
propagation vector. Various models for the magnetic ordering in \FSS{} are 
discussed in Appendix~\ref{sec:mag_models}.  The resulting magnetic structure 
shown in Fig.~\ref{fig:MagStructure} (c) and (c) is a collinear antiferromagnet 
with all moments either lying in the ab plane along (110) type directions or 
parallel to the c-axis. Because the  $(\nicefrac{1}{2},\nicefrac{1}{2},0)$ and 
$(0,0,1)$ type magnetic Bragg peaks are  coincident within the resolution of 
our measurement, the  moment direction cannot be uniquely determined from the 
powder diffraction data.  Nevertheless, refinements for each model give an 
identical ordered moment of $\langle m\rangle\!=\!  1.76(5)$~\mub{} at 1.6~K.  
This is only 44(1)\% of the 4~\mub{} saturation magnetization for the fully 
localized high spin state of Fe$^{2+}$ and $g\!=\!2$. The anomalously low 
ordered moment is typical of A-site spinels \cite{Roy:2013, Nair:2014} and may 
result from geometric frustration in this insulating material \cite{Lee:2008}.  
Indeed, the collinear antiferromagnetic ordering we observe is a highly 
frustrated ground state in the CT lattice of Fig.~\ref{fig:MagStructure}, only 
satisfying antiferromagnetic next nearest neighbour (NNN) exchange $J_2$.  Both 
the nearest neighbour (NN) $J_1$ and NNN inter-planar $J_2^{\prime}$ exchange 
terms compete with $J_2$, and cancel at the mean field level; resulting in a 
spin system that lowers its energy mainly through interactions within the 
tetragonal basal plane.

\begin{figure}[t!]
    \includegraphics[]{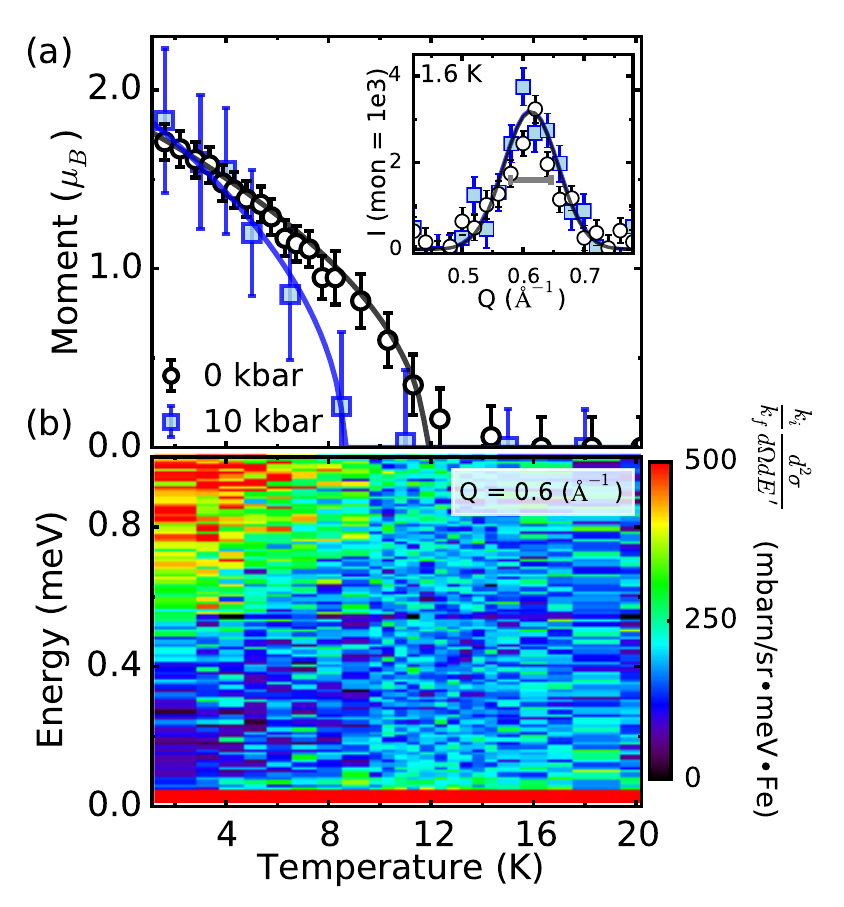}
\caption{\label{fig:OrderParam} (a) Temperature and pressure dependence of the 
    ordered magnetic moment in \FSS{}, solid lines are a fit to magnetic order 
    parameter with $\beta\!=\!  0.5$, saturated moment of 1.9(1)~$\mu_B$, and 
    T$_c$ of 11.8(2)~K and 8.6(8)~K for 0 and 10~kbar respectively .  Inset 
    shows the background subtracted magnetic Bragg peak measured under 0~kbar 
    and 10~kbar hydrostatic pressure. Horizontal line indicates instrumental 
    resolution. Error bars represent one standard deviation. (b) Temperature 
    dependent low-energy inelastic scattering measured on CNCS integrated over 
    $0.55\!<\!Q\!<\!0.65$ \AA$^{-1}$.}
\end{figure}

We now examine the thermal phase transition to this ordered state.
Fig.~\ref{fig:OrderParam} (c) shows the temperature dependent magnetic moment 
obtained from Rietveld refinements of full diffraction patterns. There is a 
smooth increase in the ordered moment upon cooling below a critical temperature 
of T$_c$=11.8(2)~K,  which at ambient pressure can be described as 
$M_0(1-T/T_c)^{\beta}$, with $\beta\!=0.5$ and $M_0\!=1.9(1)~\mu_B$.  A similar 
temperature dependence is observed under 10~kbar hydrostatic pressure, though 
with a reduced transitions temperature T$_c$=8.6(8)~K.  The inset of 
Fig.~\ref{fig:OrderParam} shows the magnetic correlation length 
($\xi\!=\!56(16)$~\AA) is unaffected by pressure to within error.

This seemingly conventional temperature dependence of the ordered moment belies 
the true nature of the magnetic phase transition as evidenced by both the 
absence of critical behaviour in the specific heat Fig.~\ref{fig:SpecificHeat}, 
and the low energy inelastic scattering at the critical wave vector of 
0.6~\AA$^{-1}$ [Fig.~\ref{fig:OrderParam} (b)]. Contrary to expectations for a 
second order phase transition, no critical scattering appears around the 
ordering wave vector at T$_c$, instead only a gradual build up of inelastic 
spectral weight, concentrated around energy transfers of $\sim1.5$~meV was 
observed.  In the frustrated pyrochlore ZnCr$_2$O$_4$, a first order transition 
abruptly selects an ordered magnetic state out of a highly degenerate manifold 
of states \cite{Lee:2000}. An inhomogeneous first order phase transition might 
relate this scenario to the smooth and latent heat free transition in \FSS{}.  

\subsection*{Magnetic Excitations}
\begin{figure*}[t!]
    \includegraphics[]{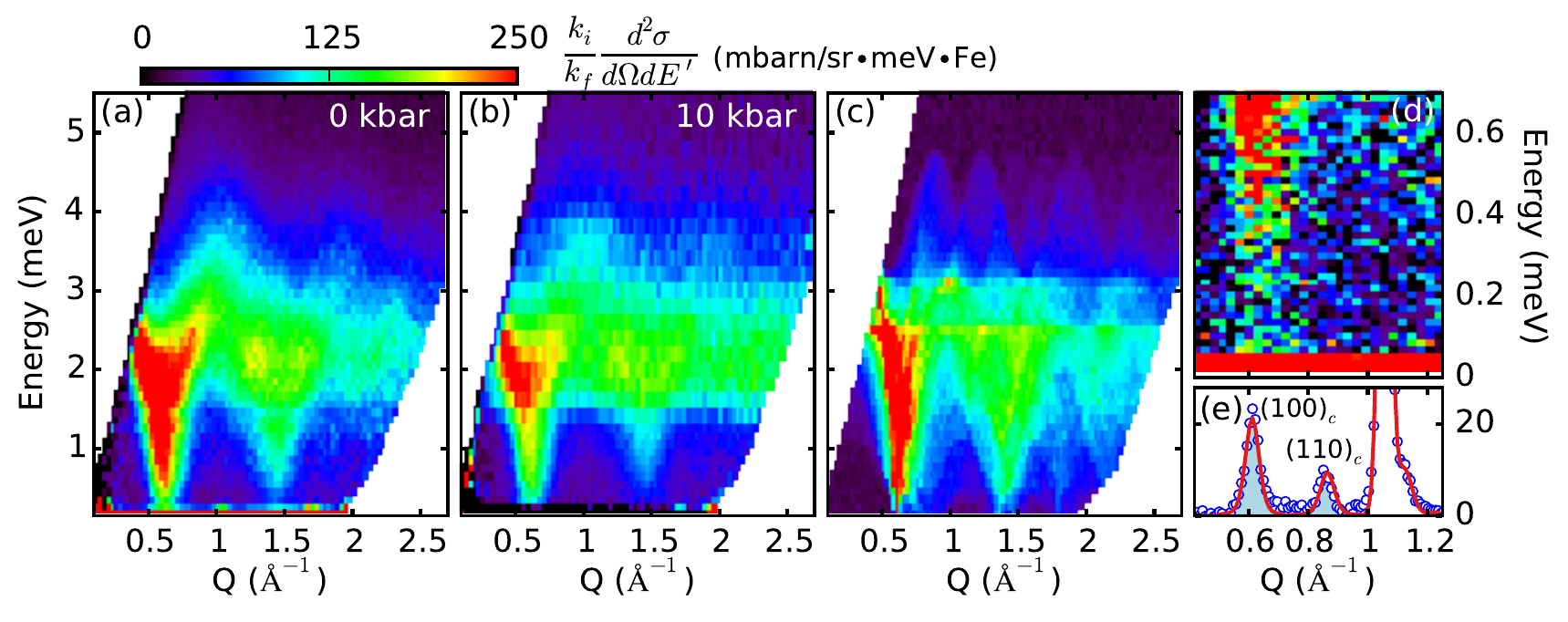}
\caption{\label{fig:SQW} (a)-(b) Powder averaged inelastic neutron scattering 
    intensity in \FSS{} measured at 1.6~K and 0~kbar in (a) and 10~kbar in (b).  
    Data has been corrected for the \FeIon form factor and placed into absolute 
    units by calibrating against the integrated intensity from a structural 
    Bragg reflection. (c) Calculated powder averaged neutron scattering 
    intensity for spin-waves in the collinear magnetic structure shown in 
    Fig.~\ref{fig:MagStructure} (b) and (c). (d) Low energy inelastic neutron 
    scattering measured on CNCS.  In panel (e) the elastic scattering is shown 
    with magnetic diffraction peaks highlighted in light blue. There is an 
    absence of inelastic intensity around the $(110)_c$ type magnetic Bragg 
    peaks.}
\end{figure*}
We now discuss the magnetic excitation spectrum in \FSS{}. An overview of the  
normalized inelastic neutron scattering intensity, $\tilde{I}(Q,E)$, is 
presented in Fig.~\ref{fig:SQW} (a) and (b) for 0~kbar and 10~kbar hydrostatic 
pressures respectively. There are intense, dispersive magnetic excitations 
emerging from Q=0.6~\AA$^{-1}$ with spectral weight extending to  
$\sim$5.5~meV.  At low energies and in the neighborhood of the critical 
wavevector (0.6~\AA$^{-1}$) the intensity continuously decreases with 
decreasing energy below $\sim$1~meV to a level just above background. One may 
be tempted to implicate a gap in the excitation spectrum based on this 
decreasing spectral weight\cite{Krimmel:2005}. However, for a
polycrystalline sample, the inelastic neutron intensity reflects the average 
intensity on the Ewald sphere, which can decrease rapidly with energy for a
dispersive, long range correlated, systems even though the excitations may be 
gapless. A gap in the excitation spectrum is signaled by the \emph{complete 
    absence} of spectral weight at low energies around the critical wavevector.  
In contrast for \FSS{} the high resolution data in Fig.~\ref{fig:SQW} (d) shows 
intensity, albeit weak, down to 0.05~meV so that energy gap in the excitation 
spectrum must be smaller than this. The excitation spectrum is consistent with 
previous reports \cite{Krimmel:2005} and, we will see, can be accounted for by 
spin wave excitations from an ordered antiferromagnet in the tetragonal cell. 

Although directional information is destroyed by powder averaging, 
$\tilde{I}(Q,E)$  holds valuable information that correlates the scalar lengths 
and energy scales. We first extract the normalized spin-space trace of the 
powder averaged dynamical spin correlation function from the neutron intensity 
$\tilde{\mathcal{S}}(Q,E) = 6 \tilde{I}(Q,E)/\left| r_0 f(Q)\right|^2$, and we 
have used the dipole approximation for the Fe$^{2+}$ ($3d^6$) form factor.  To 
determine the dominant correlation length scales independent of any model 
Hamiltonian, we compare the energy integrated inelastic signal to the 
spherically averaged equal time structure factor,
\begin{align}
    \tilde{\mathcal{S}}(Q) &=  \int_{0.2}^{6}{\!\!dE \tilde{\mathcal{S}}(Q,E)} 
    \nonumber \\
    &=\langle (s_0)^2 \rangle\!+\!  \sum_{d}\langle \rm s_0\!\cdot\!s_d \rangle  
    \frac{\sin\left(Qd\right)}{Qd},
\label{eq:static_struct}
\end{align}
where  $\langle s_0\!\cdot\!s_d\rangle$ is the average spin-spin correlator 
between sites separated by the distance $d$.  Information regarding the 
dominant magnetic bond energies may be extracted from the powder averaged first 
moment sum rule for a Heisenberg Hamiltonian 
\cite{HoenbergBrinkman:74,Stone:01},
\begin{align}
    \langle E \rangle_Q &= \int_{0.2}^{6}{\!dE E\tilde{\mathcal{S}}(Q,E)} 
    \nonumber \\
&= 2\sum_{d} B_{d} \left( 1 - \frac{\sin\left(Qd\right)}{Qd}\right) + 
\mathcal{D},
\label{eq:first_mom}
\end{align}
where $B_{d}\!=\!zJ_{d} \langle \rm s_0\!\cdot\!s_d \rangle$, is the total 
magnetic bond energy for a superexchange interaction $J$ across all bonds at 
distance $d$, $z$ is the coordination of the bond,  and  $\mathcal{D}$ is a 
constant related to the sum over all single site anisotropy energies: 
$D_{\beta}\left<\left(s_j^{\beta}\right)^2\right>$. An unknown factor of 
$\left(\nicefrac{g}{2}\right)^2$ is implicit in both eq.~\ref{eq:static_struct} 
and eq.~\ref{eq:first_mom} and is contained within the correlators and bond 
energies quoted here. 

\begin{figure}[t!]
    \includegraphics[]{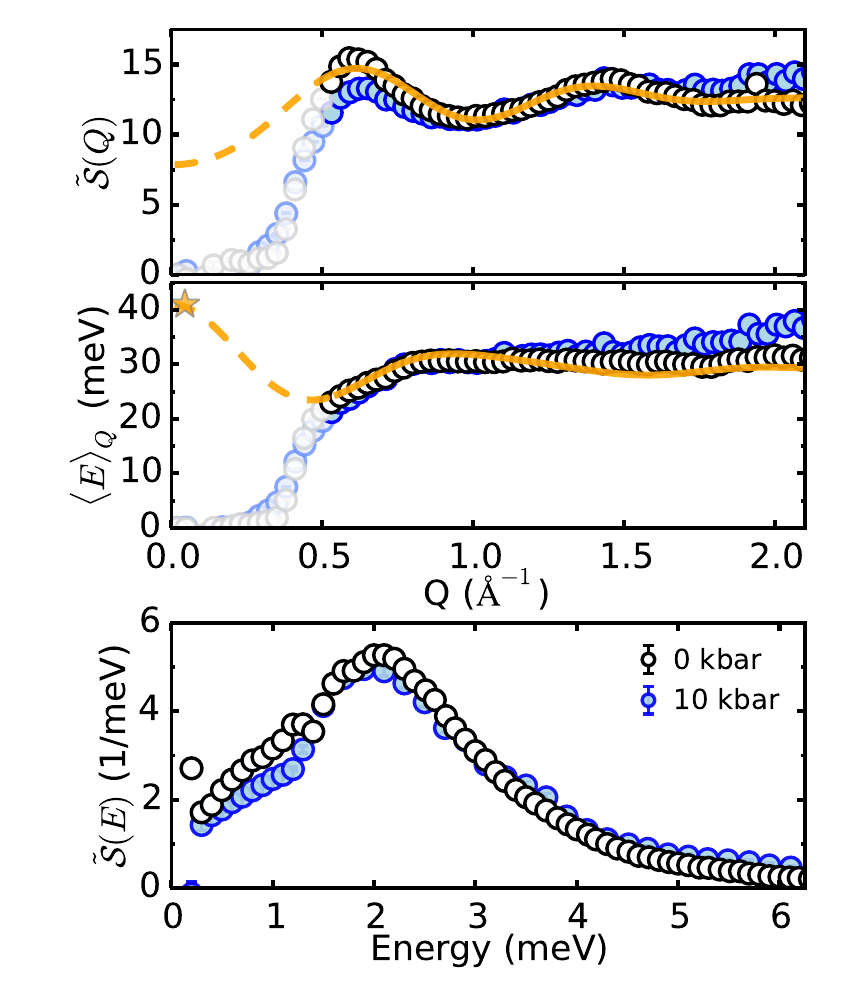}
\caption{\label{fig:Cuts} (a) Static structure factor 
    $\tilde{\mathcal{S}}\left(Q\right)$ determined by integrating the measured 
    intensity between $0.2\!<\!E\!<\!6$~meV and correcting for the Fe$^{2+}$ 
    form factor.  Solid line is a fit to a powder averaged structure factor as 
    described in the text.  (b) First moment of the neutron scattering 
    intensity integrated between $0.2\!<\!E\!<\!6$~meV. The yellow star 
    indicates the Q=0 first moment as extrapolated from Terahertz measurements 
    and the static structure factor.  Solid line is a fit to the powder 
    averaged first moment sum-rule described in the text.  In (a) and (b) the 
    dashed line spans the region not included in the fit because of kinematic 
    limitations.  (c) Momentum integrated inelastic scattering, integrated 
    between $0.45\!<\!Q\!<\!1.6~$\AA$^{-1}$. Error bars representing one 
    standard deviation are smaller than the symbol size.
    }
\end{figure}
The energy integrated inelastic intensity and first moment are shown in 
Fig.~\ref{fig:Cuts} (a) and (b) respectively. Due to the kinematic limit for 
inelastic neutron scattering, the data only sample the full spectrum  for 
$Q\!>\!0.5$~\AA$^{-1}$.  While the static structure factor exhibits well 
defined peaks distinguishing length scales that dominate the magnetic 
correlations, the first moment is relatively featureless and flat as a function 
of momentum transfer. A momentum independent first moment  implies the ground 
state energy is dominated either by a local energy scale through $\mathcal{D}$ 
or by potentially frustrated interactions covering a wide range of length 
scales.  The two contributions are clearly distinguished in the Q=0 limit where 
magnetic bond energy terms in equation~\ref{eq:first_mom} vanish.  Because of 
the kinematic limits for inelastic neutron scattering, a fit of 
equation~\ref{eq:first_mom} to the neutron scattering data alone is 
insufficient to constrain $\mathcal{D}$.  However, at Q=0 the first moment and 
static structure factor are simply related $\langle E \rangle_{Q=0} = 
\mathcal{D} = \langle \hbar\omega\rangle_{0}\tilde{\mathcal{S}}(0)$ where 
$\langle\hbar\omega\rangle_{0}$ is the energy averaged over the fluctuation 
spectrum at Q=0. Time domain Terahertz spectroscopy has revealed a single well 
defined mode at $\hbar\omega\!=\!4.5$~meV which dominates the dynamic 
susceptibility at $Q\!=\!0$ \cite{Laurita:2015}; using this result we may apply 
a single mode approximation to obtain a lower bound on the anisotropy energy 
$\mathcal{D}=4.5\tilde{\mathcal{S}}(Q=0)$.  Fits to the static structure factor 
and first moment are shown in Fig.~\ref{fig:Cuts} and the extracted correlators 
and bond energies are listed in Table~\ref{tab:bond_params}. 

From the static structure factor fit we find $\langle(s_0)^2\rangle\!=\!13(1)$, 
and  $\mathcal{D}\!  =\!  40(5)~{\rm meV}$.  Given their similar ranges, 
magnetic interactions labelled $J_i$ and $J_i^{\prime}$ in 
Fig.~\ref{fig:MagStructure} (b) and (c) cannot be distinguished and so are 
represented by their average.  Any differences in the magnetic correlators and 
bond energies between the 0~kbar and 10~kbar data are not discernible from the 
effects of increased background scattering resulting from the stainless steel 
pressure vessel. We conclude that hydrostatic pressures of 10~kbar has little 
effect on the overall magnetic energy scales, though T$_{\rm c}$ is reduced 
slightly and there is some suppression of low energy scattering 
[Fig.~\ref{fig:SQW} (a) and (b)].
\begin{table}[h]
\caption{Exchange interactions and spin-spin correlators across magnetic bonds 
    in \FSS{}. Bonds are labeled according to Fig. \ref{fig:MagStructure} (b) 
    and (c). The correlators and bond energies are extracted from fits to 
    experimental data while exchange interactions $J_{sw}$ are those used for 
    the spin wave model renormalized by the reduced ordered moment 
    $\frac{\langle m \rangle}{gS}\!=\!\frac{1.76}{4}$.}
\begin{tabular}{ccccc}
\hline
\hline
 & d (\AA) & $\langle s_0\!\cdot\!s_d\rangle$  & $B_d$ (meV)  & 
 $\nicefrac{\langle m \rangle}{(gS)} \cdot J_{sw}$ (meV) \\
 \hline
$J_1$           & 4.55  &  1(1)    &   3.7(7)     & -0.14(8)  \\
$J_2$           & 7.43  & -7(2)   &  -4.1(7)    & 0.7(1)   \\
$J_2^{\prime}$  & 7.43  & --        &   --        & 0.6(6)   \\
$J_3$           & 8.70  & -2(1)   &  -2.3(3)     &  0.02(2)  \\
$J_3^{\prime}$  & 8.72  & --        & --        & 0.02(2)   \\
$J_4$           & 10.51 & 3(1)    & -3.1(3)     & -0.09(7)  \\
$J_4^{\prime}$  & 10.49 & --        & --        &-0.09(7)   \\
\hline
\hline
\end{tabular}
\label{tab:bond_params}
\end{table}

The sign and magnitude of spin correlators we extracted are entirely consistent 
with the collinear magnetic order shown in Fig.~\ref{fig:MagStructure} (a) and 
support a  model where dominant antiferromagnetic correlations exist across the 
NNN bonds ($J_2$) of the Fe lattice. This analysis suggests that, at least in 
terms of the magnetic correlations, the system may be considered as essentially 
two interpenetrating CT sub-lattices, where $J_2$ couples spins on the same 
sublattice and $J_1$ fails to produce significant correlations. Interestingly, 
$\mathcal{D}$ emerges as the dominant energy scale.  This is indicative of a 
magnetic state which breaks the cubic symmetry as all single ion anisotropy 
terms must vanish in the cubic cell. However, $\mathcal{D}$ is anticipated for 
an orbitally ordered state in the presence of spin orbit interactions.  
Furthermore, such a relatively large single-ion anisotropy scale reveals an 
instability of the magnetically ordered state towards a local singlet.      

Fig.~\ref{fig:Cuts} (c) shows the momentum averaged inelastic intensity which 
approximates the magnetic density of states.  The effect of 10~kbar
hydrostatic pressure is a slight transfer of spectral weight from around 1~meV 
to energies between 5 and 6~meV. 

Integrating the total measured dynamical spin correlation function over the 
region $0.2\!<\!E\!<6$~meV and $0.45\!<\!Q\!<\!1.6$~\AA$^{-1}$, we obtain the 
total inelastic spectral weight $\langle \delta m^2 \rangle\!=\!\int\!\int\!Q^2 
\tilde{\mathcal{S}}(Q,E) {\rm d}Q{\rm d}E/\!\int\!  Q^2{\rm d}Q$, recovering a 
fluctuating moment of  $\langle \delta m^2 \rangle \!=\!13(1)$ at 0~kbar and  
$\langle \delta m^2 \rangle \!=\!12(1)$ at 10~kbar.  Summing the static 
(elastic) and dynamic (inelastic) contributions yields a total moment of 
$m^2_{tot}\!=\!\langle m \rangle^2 \!+\!  \langle \delta m^2 \rangle 
\!=\!16(3)$, significantly less than $g^2S\left(S+1\right)\!=\!24$ expected for 
S=2 and g=2.  There are a number of not mutually exclusive explanations for 
this discrepancy: (i) Significant inelastic spectral weight exists beyond the 
6~meV range of the experiment. (ii) There is a component of elastic diffuse 
magnetic scattering that arises from static disorder.  The second scenario is 
difficult to test with our un-polarized measurement as elastic  diffuse 
scattering is dominated by incoherent nuclear scattering from Sc 
($\sigma_{inc}\!=\!4.5$~barn) and any elastic diffuse magnetic scattering will 
be broad in momentum and extremely weak.  However, we do not observe any 
temperature or momentum dependent background which may arise from magnetic 
diffuse scattering below 20~K and the first scenario thus seems more likely 
based on our data. This conclusion is supported by our specific heat 
measurement which only recover 70~\% of the available magnetic entropy by 
$\sim$70~K (6~meV), consistent with higher energy magnetic spectral weight.  
Surprisingly, for such a large spin (S=2) antiferromagnet, the dynamic portion 
dominates the spectral weight, contributing over four times that of the static 
portion. For a fully spin polarized ground state, the magnetic inelastic 
scattering consists entirely of spin excitations which are transverse to the 
ordered moment direction (spin waves) and have an available spectral weight of 
$g^2S(S+1)\!-\!g^2S^2\!=g^2S\!\approx\!8$, when $g$=2.  Quantum fluctuations, 
which reduce the ordered moment, allow for spin excitations which are 
longitudinal fluctuations and an associated increase of inelastic spectral 
weight. It then follows that the large fluctuating moment we observe implies 
the N\'eel state in \FSS{} is strongly renormalized by quantum fluctuations 
which may result of disorder and/or an incipient quantum  phase transition.

Having identified pertinent magnetic interaction pathways in \FSS{} we may now 
compare the measured excitation spectrum with a minimal effective spin 
Heisenberg model. While one may contest the validity of a spin wave expansion 
for such a strongly renormalized state as in \FSS{} with  $\langle\left(\delta 
    m\right)^2\rangle\!=\!13>g^2S\!=\!8$, the semi-classical description might 
provide a reasonable account of long-wave-length dispersive transverse 
components of the magnetic excitation spectrum.  Figures~\ref{fig:SQW} (d) and 
(e) juxtapose the low-energy inelastic neutron scattering signal with the 
elastic signal in the neighborhood of the first two magnetic Bragg peaks, 
indexed in the cubic unit cell as (100)$_c$ and (110)$_c$ respectively.  
Comparison of the elastic and inelastic intensities in this region immediately 
reveals an important clue that proves essential to determine the magnetic order 
in \FSS{}.  While there is strong elastic scattering both around (100)$_c$ and 
(110)$_c$, inelastic scattering is found only around (100)$_c$ 
(0.6~\AA$^{-1}$).  Since (100)$_c$ and (110)$_c$ are equivalent by symmetry, 
the energy eigenvalues of magnetic excitations must be identical at these wave 
vectors.  The absence of powder averaged inelastic scattering near (110)$_c$ 
must then reflect different matrix elements for $Q\!=\!0.6$~\AA$^{-1}$ and 
0.84~\AA$^{-1}$. In the low energy long-wave-length limit the inelastic 
magnetic intensity $\tilde{I}(Q,E)$ scales with the elastic scattering except 
for the polarization factor, which extinguishes scattering associated with spin 
components along wave vector transfer, $\mathbf{Q}$. For magnetic structures 
that are consistent with the observed diffraction pattern in the cubic cell, 
the polarization factor can only at most extinguish half of the spectral weight 
at (110) compared to (100).  The suppression observed experimentally is 
however, more than a factor of 4 so that indexing of the magnetic structure 
with a cubic unit cell must be rejected. It is this feature of the inelastic 
scattering which implicates the
$q_m\!=\!(\nicefrac{1}{2},\nicefrac{1}{2},0)$ magnetic ordering proposed in 
figure~\ref{fig:MagStructure} where the two peaks correspond to distinct points 
in the Brillouin zone, $(\nicefrac{1}{2},\nicefrac{1}{2},0)$ and $ 
(\nicefrac{1}{2},\nicefrac{1}{2},1)$, that can have different excitation 
spectra. We shall now show that spin waves in a 
$q_m\!=\!(\nicefrac{1}{2},\nicefrac{1}{2},0)$ structure can account for these 
unusual low energy features of \FSS{} and indeed for the qualitative features 
of the full inelastic magnetic neutron scattering cross section. 

Within the tetragonal structure there are at least seven magnetic exchange 
pathways identified in Fig.~\ref{fig:MagStructure} (b) and (c) that must be 
considered in a simple spin wave model.  Additionally, the two Fe$^{2+}$ sites 
in the unit cell have distinct tetragonal environments, permitting two distinct 
single ion anisotropy terms in the magnetic Hamiltonian. Dzyaloshinksii 
exchange terms are also symmetry allowed on each of the magnetic exchange paths 
in the tetragonal cell. The powder averaged inelastic magnetic neutron 
scattering cross section was computed from a minimal Hamiltonian and linear 
spin wave theory (LSWT) as implemented in the \texttt{SpinW} program 
\cite{Toth:spinw}. For simplicity we include only Heisenberg terms which are 
labelled in Fig.~\ref{fig:MagStructure} (b) and (c) plus distinct single-ion 
easy plane anisotropy terms on each Fe$^{2+}$ site. Guided by the static 
structure factor and first moment sum rule analysis, we  adjusted the exchange 
constants to obtain the best qualitative agreement between the measured and 
calculated spectra.  The results of this calculation are shown in 
Fig.~\ref{fig:SQW} (c) for the exchange interactions detailed in 
table~\ref{tab:bond_params} and single ion easy plane anisotropy energies 
$D_{ab}\!=\!0.10(5)$~meV and $D_{ab}^{\prime}\!=\!0.05(4)$~meV. To aid 
meaningful comparison of the LSWT derived exchange constants and magnetic bond 
energies in the strongly fluctuating N\'eel state of \FSS{}, LSWT exchange 
constants in table~\ref{tab:bond_params} have been renormalized by the reduced 
ordered moment $\nicefrac{\langle m \rangle}{gS}\!=\!\nicefrac{1.76}{4}$. The 
relative magnitudes of exchange interactions from the LSWT are in reasonable 
agreement with those extracted from the sum-rule analysis and it is encouraging 
that the magnitude of these exchange parameters are generally consistent with 
band structure calculations for \FSS{} \cite{Sarkar:2010}.

Although the powder averaged neutron scattering data does not contain 
sufficient information to adequately constrain the seven exchange and two 
anisotropy parameters. The purpose of the spin wave calculation presented here 
is not to determine the ground state Hamiltonian, but to show that the  
character particularly of low energy spin excitations in \FSS{} can largely be 
explained in terms of a spin wave model. The spin wave model captures the 
general intensity distribution including the concentration of high energy 
spectral weight around Q$\approx$ 1~\AA$^{-1}$ and the inelastic intensity 
differences around the $(\nicefrac{1}{2},\nicefrac{1}{2},0)$ and $ 
(\nicefrac{1}{2},\nicefrac{1}{2},1)$ magnetic reflections. In fact, the 
excitation spectrum is only consistent with a magnetic propagation vector of 
$q_m\!=\!(\nicefrac{1}{2},\nicefrac{1}{2},0)$ in the tetragonal cell and the 
particular magnetic ordering pattern shown in Fig.~\ref{fig:MagStructure} (b).  
A comparison of the calculated linear spin wave neutron intensity for different 
models of magnetic ordering in \FSS{} are presented in 
Appendix~\ref{sec:spinwaves}. The  weak scattering intensity around 
$(\nicefrac{1}{2},\nicefrac{1}{2},1)$  is a direct result of the highly 
frustrated nature of the magnetic ordering in the tetragonal cell. For the 
collinear ordering, the dominant exchange interactions which have a component 
along the c-axis ($J_2^{\prime}$ and $J_1$) are frustrated and cancel at the 
mean field level.  This effectively results in dimensional reduction through 
frustration.  One consequence of this frustration is softening of magnetic 
excitations around $(\nicefrac{1}{2},\nicefrac{1}{2},1)$, which manifests as 
weak diffuse scattering intensity near 0.85~\AA$^{-1}$ in the powder averaged 
spectrum. 

Comparison with the spin wave expansion yields a further important insight. The 
inclusion of \emph{easy plane} anisotropy terms are essential to stabilize the 
magnetic structure for agreement between model and data. These terms only gap 
transverse excitations which are spin fluctuations out of the a-b plane. At the 
level of linear spin wave theory, there is always a gapless transverse 
fluctuation within the a-b plane.  Thus, inasmuch as the spin wave expansion 
accurately accounts for magnetic excitations in \FSS{}, the data are consistent 
with a gapless excitation spectrum. 

While the spin wave model appears to capture most features of the excitation 
spectrum we do not claim it to be unique. Furthermore, it is  not
clear that a spin wave expansion should be relevant for the highly renormalized 
N\'{e}el state of \FSS{}. Discrepancies of the spin wave model are 
particularly apparent at higher energies, above 2~meV, where longitudinal and 
multimagnon excitations that are not accounted for in the $\nicefrac{1}{S}$ 
expansion may be expected to play an important role.  The spectral weight 
available for two-magnon (longitudinal) scattering $\langle \delta 
m_{2M}^2\rangle = \Delta S \left(\Delta S+1\right)$ is directly related to the 
reduction in sublattice magnetization $\langle m \rangle^2 = \left(S-\Delta 
    S\right)^2$,  and $\Delta S \!=\!  S - \langle S_z \rangle$ .  Assuming the 
moment reduction is entirely a consequence of quantum fluctuations, the total 
fraction of spectral weight observed which results from non-spin wave 
excitations is   $\Delta S \left(\Delta S+1\right)/S\left(S+1\right)\!=\!  
\nicefrac{2.391}{6}=40\%$.

\section*{Discussion}
We have reported long range magnetic ordering in \FSS{} which appears at a 
temperature concomitant with a broad turnover in the specific heat.  We have 
presented evidence that this order arises from a centered tetragonal space 
group where the single ion site associated with cubic symmetry is split into 
two orbitally non-degenerate sites.  Static and dynamic magnetic correlations 
find a consistent explanation within a tetragonal unit cell and with a magnetic 
propagation vector $q_m\!=\!(\nicefrac{1}{2},\nicefrac{1}{2},0)$.  Although the 
unit cell remains  metrically cubic, the alternating compression and expansion 
of the coordinating sulfur tetrahedra lifts the orbital degeneracy of the ideal 
tetrahedral coordination. In concordance with these results we find the 
magnetic ground state energy contains significant contributions from 
anisotropic terms in the Hamiltonian that are incompatible with an orbitally 
degenerate cubic structure.  Together, these observations show the ground state 
of \FSS{} is not a spin orbital liquid as previously hypothesized. 

The upper bound we may place on a tetragonal distortion in \FSS{} is extremely 
small, 0.2\%, and spin orbit coupling may still enter as an energy scale 
competitive with the small non-cubic crystal field. In this case, we may 
naively apply the theory developed for the spin-orbital liquid phase to  
estimate the proximity of \FSS{} to the quantum critical point separating the 
spin-orbital liquid phase from a long range ordered phase.  In the mean field 
phase diagram proposed by Chen \emph{et  al.} \cite{Chen:2009} the critical 
point occurs at $x_c\!=\!\nicefrac{J_2}{\lambda}\!=\!\nicefrac{1}{16}$.  Using 
the experimentally determined value for the SOC coupling constant 
$\lambda\!=\!1.57(25)$ meV \cite{Laurita:2015} and the unrenormalized magnetic 
exchange $J_2$ used in the spin wave model we find 
$\nicefrac{J_2}{\lambda}\!>\!0.20$ which falls well within the magnetic and 
orbitally ordered regime. We note that the Terahertz measurements also extract 
a value of  $\nicefrac{J_2}{\lambda}\!>\!\nicefrac{1}{16}$, but based on the 
reported absence of magnetic order in previous neutron measurements conclude 
that quantum fluctuations renormalize $x_c$ \cite{Laurita:2015}. The 
drastically reduced  ordered moment of \FSS{} and enhanced fluctuations we 
observe are a direct sign of the  melting of the staggered magnetization in 
proximity to a critical point. Furthermore, the dominant single ion anisotropy 
energy scale we extract indicates that this instability is driving towards a 
local singlet state, this is precisely the nature of the spin-orbital singlet 
state.  Given the proximity, it may be possible to drive \FSS{} through the 
quantum phase transition and stabilize the spin orbital liquid ground state by 
application of a suitable perturbation \cite{Chen:2009b, Ish:2015}.  We have 
found a reduced staggered magnetization and shift of spectral weight towards 
higher energies under a hydrostatic pressure of 10~kbar. Perhaps it is possible 
to reach the critical point through application of even higher pressure. When 
single crystals become available uniaxial stress is another interesting 
parameter that may act to drive \FSS{} either deeper into the ordered state (as 
indicated by $\nicefrac{c}{a}\!=\!0.998(2)$) or towards the critical point. 

The nature of the thermal magnetic phase transition leading to this strongly 
fluctuating N\'eel state is rather peculiar.  Specifically, how can the 
apparent critical behaviour of the magnetic order parameter be reconciled with 
the broad turnover in specific heat and lack of magnetic critical scattering?  
There are a number of possibilities.  One simple explanation is that the sample 
is inhomogeneous and only a small fraction it orders, perhaps as a result of 
random impurities. The diamond lattice antiferromagnet is particularly 
sensitive to this through the order by quenched disorder mechanism 
\cite{Savary:2011}.  A partial magnetic order induced by quenched disorder 
would be consistent with the small ordered moment we observe since the neutron 
signal is sensitive only to the product of the ordered moment and volume 
fraction of the ordered phase.  

This is contradicted by the fact that the inelastic magnetic scattering can be 
described by spin waves in an ordered, renormalized, antiferromagnet.  Our 
samples also appear cleaner than those typically reported in literature where 
magnetic ordering has not been reported.  Structural refinements at 100~K 
definitively exclude any non-stoichiometries beyond a $\sim$1\% limit of 
detection, and also tightly constrain any Fe:Sc mixing to less than 3\% (See 
Appendix~\ref{sec:S_XRD}).  Comparison of thermodynamic measurements on our 
samples and those of previous reports further support this conclusion. The low 
temperature maximum our measured specific heat both occurs at a higher 
temperature and is sharper in temperature than previously reported 
\cite{Fritsch:2004}. Bulk magnetic susceptibility of our powder samples, 
discussed in Section~\ref{sec:suscept}, show a turnover at $\sim$10~K closely 
resembling the local susceptibility obtained from NMR Knight shift measurements 
\cite{Buttgen:2004,Buttgen:2006}.  This is a clear indication the absence of 
magnetic impurities or orphan spins in our samples.  These typically lead to a 
low temperature Curie tail that was observed in previous reports though not in 
our sample \cite{Fritsch:2004,Buttgen:2004}.  Moreover, we have observed the 
magnetic Bragg reflections in three different samples each having undergone a 
different thermal processing schedule. While the maximum temperature during 
synthesis differed by as much as 400~$^{\circ}$C between samples, the  relative 
intensity of the $\left(\nicefrac{1}{2},\nicefrac{1}{2},0\right)$
(Q=0.6~\AA$^{-1}$) magnetic reflection and (1,1,1) (Q=1.03~\AA$^{-1}$)  
structural Bragg peak vary by no more than a factor of $\nicefrac{1}{3}$.

The behaviour observed in \FSS{} is similar to another A-site magnetic spinel 
CoAl$_2$O$_4$ \cite{Roy:2013},  where neutron scattering measurements on single 
crystals show signatures of antiferromagnetic domains that have been argued to 
result from an arrested first order, order-by-disorder  magnetic transition 
\cite{MacDougall:2011}.  The 50~\AA{} correlation length we observe in \FSS{} 
is consistent with the presence of large antiferromagnetic domains.  
Furthermore, Monte Carlo simulations of the order-by-disorder transition in the  
face centered cubic lattice show that thermally excited defects and domain 
walls drastically reduce the sublattice magnetization of the ordered state 
\cite{Gvordikova:2005}. The collinear ordering we observe in \FSS{} is only 
very weakly favored over other possible magnetic structures.  A classical 
order-by-disorder mechanism should, however, generally favor a collinear 
magnetic ordering \cite{Henley:1989}.  Indeed, both random strains as well as 
spin-orbit coupling have been invoked to explain the quadrupolar splitting 
observed by M\"ossbauer experiments on \FSS{} \cite{Brossard:1976}.  The 
structural distortion we observe may hint at another mechanism: in the 
tetragonal cell there is an alternating compression and elongation of the Fe 
coordinating S tetrahedron, but the diffraction measurement does not uniquely 
identify which iron site has the compressed and which elongated tetrahedral 
environment. If the tetrahedral distortion is not coherent between many unit 
cells the result will be a static orbital disorder and as a consequence 
exchange disorder. Since the structural unit cell remains constant, the thermal 
energy barrier to creating such a magnetic exchange defect will be very low.

In this order-by-disorder scenario, an abrupt first order transition becomes 
rounded by the presence of disorder \cite{Imry:1979, Goswami:2008}. Rather than 
a single sharp transition, the exchange disorder results in a distribution of 
first order transition temperatures whereby  larger and larger volumes of the 
sample consecutively undergo abrupt first order transitions upon cooling. The 
outcome appears as a gradual onset of order when in fact all volume elements of 
the sample actually undergo an abrupt first order transition \cite{Hui:1989}.  

Even though our experiment now reveals it to lie on the opposite side of the 
critical point as was previously anticipated, \FSS{} appears as a unique 
material where many energy scales including crystal field splitting, spin-orbit 
coupling, and frustrated magnetic exchange interactions compete and conspire to 
produce very rich physics. This includes a thermal order-by-disorder transition 
into a highly renormalized N\'eel ground state on the border of a quantum phase 
transition.

\begin{acknowledgments}
We would like to thank N. P. Armitage and N.J. Laurita for helpful discussions 
and comments on this manuscript.  We also acknowledge Juscelino Leao for 
assistance with the high pressure sample environment and Nicholas Butch for 
providing transmission measurements of the stainless steel pressure vessel.  
Work at IQM was supported by the U.S.  Department of Energy, Office of Basic 
Energy Sciences, Division of Material Sciences and Engineering under grant 
DE-FG02-08ER46544. Work at ORNL was sponsored by the Division of Scientific 
User Facilities, Office of Basic Energy Science, US department of Energy (DOE).  
Work at NIST utilized facilities supported in part by the National Science 
Foundation under Agreement No.  DMR-1508249.
\end{acknowledgments}

\appendix
\setcounter{equation}{0}
\renewcommand{\theequation}{A\arabic{equation}}
\section{Sample Characterization}
\subsection{Synchrotron Powder Diffraction}
\label{sec:S_XRD}
\begin{figure}[h!]
    \includegraphics[]{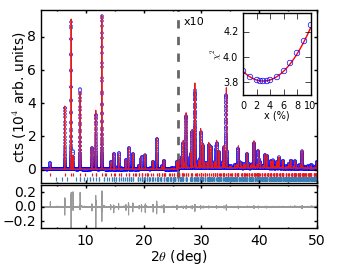}
\caption{\label{fig:Structure_refine} Rietveld analysis of high resolution 
    synchrotron x-ray data collected on 11-BM at 100~K.  Blue circles are 
    measured data, red line is the calculated diffraction pattern and red 
    vertical bars indicate the positions of nuclear reflections in the \Fdm{} 
    space group. There is also a contribution from a 0.15 \% Sc$_2$O$_3$ 
    contamination phase. The difference between calculated and measured data is 
    shown in the lower panel. Inset shows the goodness of fit parameter 
    ($\chi^2$) versus the Fe-Sc anti-site inversion. The Sc$_2$O$_3$ 
    contamination phase was not included in refinements used to determine the 
    site-inversion.}
\end{figure}
The $T\!=\!100$~K synchrotron powder diffraction data pattern collected on 
11-BM and refinement is shown in figure~\ref{fig:Structure_refine}. All 
diffraction peaks are well indexed using the cubic \Fdm{} unit cell and we find 
a good  refinement to the 100~K diffraction data using the reported structure 
for \FSS{}. However, significant microstrain broadening of the diffraction 
peaks was necessary to accurately describe the peak shapes.  An isotropic 
strain broadening in the cubic cell was accounted for in GSAS utilizing the 
semi-empirical Stephens peak shape \cite{Stephens:1999} with  cubic strains of  
S$_{400}$= 0.19\%, S$_{200}$ = 0.01\%. The large difference  between the two 
symmetry allowed strain parameters is consistent with an incipient structural 
transition.  Our x-ray powder diffraction data additionally show the presence 
of a minor (0.15 \%) contamination phase of Sc$_2$O$_3$ in our sample, below 
the limit of detection for our neutron diffraction measurements.

The magnetic properties of A-site spinel compounds are highly sensitive to the 
degree of chemical disorder resulting from inversion between atomic species 
occupying the $A$ and $B$ sites of the spinel structure \cite{Hanashima:2013, 
    Roy:2013}. To quantify any inversion present in our sample we have refined 
out synchrotron data allowing for anti-site disorder as described by 
(Fe$_{1-x}$Sc$_x$)[Sc$_{2-x}$Fe$_x$]S$_4$. The inset of 
figure~\ref{fig:Structure_refine} shows the goodness-of-fit parameter versus 
the site inversion $x$ which indicates a small amount of site-inversion 
$x=0.03(3)$ present in our sample. The value of $\chi^2$ in the inset of 
figure~\ref{fig:Structure_refine} is larger than reported for the full Rietveld 
refinement because only a single phase was included in Rietveld refinements 
used to estimate the site inversion. We have checked that the inclusion of the 
contamination phase does not effect the conclusion of the site mixing analysis.  
Rietveld refinement parameters are listed in Table~\ref{tab:HT_struct}.

\begin{table}[t!]
\caption{Atomic parameters obtained from Reitveld refinements of synchrotron 
    x-ray diffraction data at 100~K for \FSS{} samples used in this work. The 
    space group is $F\bar{d}32$ (227) and lattice parameter was refined to 
    $a\!=\!10.51115(1)$~\AA.  Reitveld refinements resulted in a $\chi^2$ of 
    2.83 and $R_{Bragg}\!=\!5.59$\%.}
\begin{tabular}{lcccccc}
\hline
\hline
Atom & Site & x & y & z & Occ &B$_{iso}$\\
\hline
Fe & 8a & 0.125   & 0.125 & 0.125 & 0.94  & 0.313(2) \\
Sc & 8a & 0.125   & 0.125 & 0.125 & 0.06  & 0.311(2) \\
Fe & 16d  & 0.500   & 0.500 & 0.500 & 0.03 & 0.313(2) \\
Sc & 16d  & 0.500   & 0.500 & 0.500 & 0.97 & 0.311(2) \\
S & 32e  & 0.2556(1)   & 0.2556(1) & 0.2556(1) & 1.00  & 0.352(2) \\
\hline
\hline
\end{tabular}
\label{tab:HT_struct}
\end{table}

\subsection{Bulk Magnetization}
\label{sec:suscept}
\begin{figure}[t!]
    \includegraphics[]{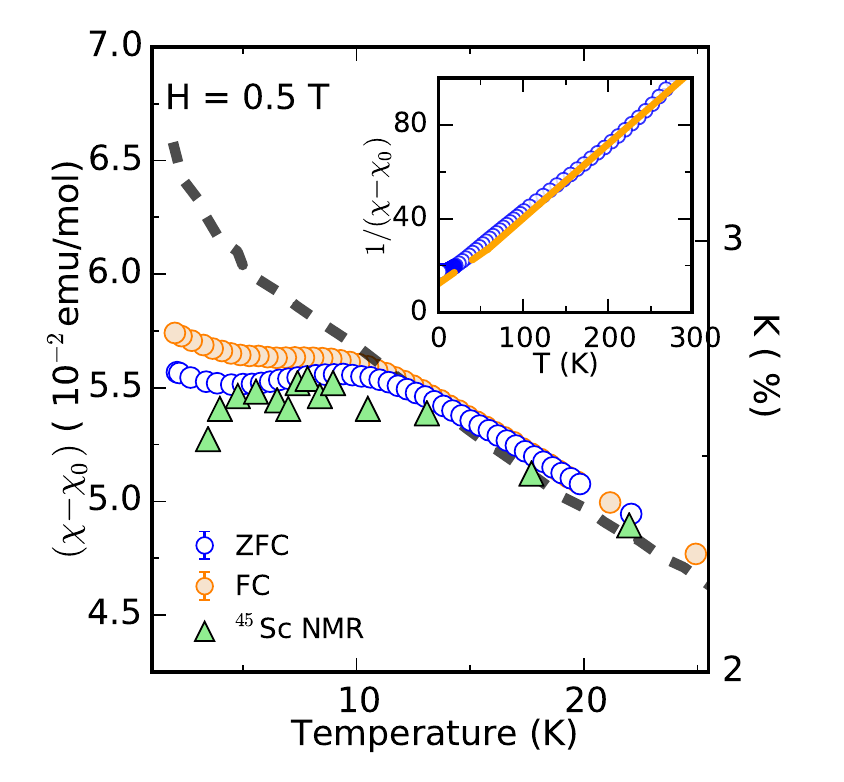}
    \caption{\label{fig:suscept} Magnetic susceptibility of \FSS{}. The low 
        temperature susceptibility is shown in the main panel. Open circles are 
        the susceptibility measured on our sample, dashed line is bulk 
        susceptibility data from Fritsch \emph{et al.} \cite{Fritsch:2004}, and 
        triangles show the NMR Knight shift data from \cite{Buttgen:2004}.  
        Inset shows the inverse susceptibility solid line is a linear fit to a 
        Curie-Weiss law.}
\end{figure}

Temperature dependent bulk magnetization of \FSS{} was measured using a Quantum 
Designs PPMS in an applied field of 0.~T and is shown in 
figure~\ref{fig:suscept}.  The high temperature susceptibility displays 
Curie-Weiss behavior with a paramagnetic moment of 5.1(1) \mub{} and 
Curie-Weiss temperature $\theta_{CW}\!=\!-40.1(5)$ K, consistent with earlier 
reports \cite{Fritsch:2004}. The effective moment for $g\!=\!2$ and $S\!=\!2$ 
is $g\sqrt{S(S+1)}\!=\!4.90$.
At lower temperatures our data deviate from previous reports.  The data show a 
broad turnover at $\sim$10~K, where reports in the literature typically have a 
5\% or more Curie tail \cite{Fritsch:2004}. This turnover is identical to the  
reported local susceptibility from Sc NMR Knight shift measurements 
\cite{Buttgen:2004,Buttgen:2006}. The NMR Knight shift probes the intrinsic 
local spin susceptibility, independent of any Curie tail contribution from 
small amounts of impurities which may dominate the bulk susceptibility at low 
temperatures. Agreement between the measured bulk susceptibility of our samples 
and NMR Knight shift measurements is a  strong indication of the absence of 
magnetic impurities which are present in previously reported samples of \FSS{}.

\subsection{Phonon contribution to Specific Heat}
\label{sec:phonon_cp}
Phonon contributions to the specific heat of \FSS{} were estimated by scaling 
the measured specific heat of the non-magnetic analog spinel compound 
CdIn$_2$S$_4$.  The measured specific heat for CdIn$_2$S$_4$ was extracted from 
\cite{Fritsch:2004}, in order to describe the phonon contribution to \FSS{}, 
this must be scaled by the relative Debye temperatures, $\Theta$,  of the two 
compounds,
\begin{equation}
  \frac{\Theta^3_{CdIn_2S_4}}{ \Theta^3_{FeSc_2S_4}} = \left(\frac{m_{Fe} + 
          2m_{Sc} + 4m_{S}}{m_{Cd} + 2m_{In} + 4m_{S}}\right)^{\nicefrac{3}{2}} 
  ,
\end{equation}
where $m_x$ is the atomic mass of element $x$. The rescaled temperature for the 
phonon specific heat is then given by
\begin{equation}
    T_{FeSc_2S_4} = \frac{\Theta_{CdIn_2S_4}}{\Theta_{FeSc_2S_4}}\cdot 
    T_{CdIn_2S_4},
\end{equation}
and the rescaled lattice contribution by
\begin{equation}
    C_{FeSc_2S_4} = \frac{\Theta_{CdIn_2S_4}^3}{\Theta_{FeSc_2S_4}^3}\cdot
    C_{CdIn_2S_4}.
\end{equation}
The scaled phonon contribution to the specific heat of \FSS{} is shown as a red 
dashed line in figure~\ref{fig:SpecificHeat} of the main text.

\section{Structural and Magnetic Models}
\subsection{Low Temperature Structural Refinements}
\label{sec:struct_models}
\begin{figure*}[ht!]
    \includegraphics[]{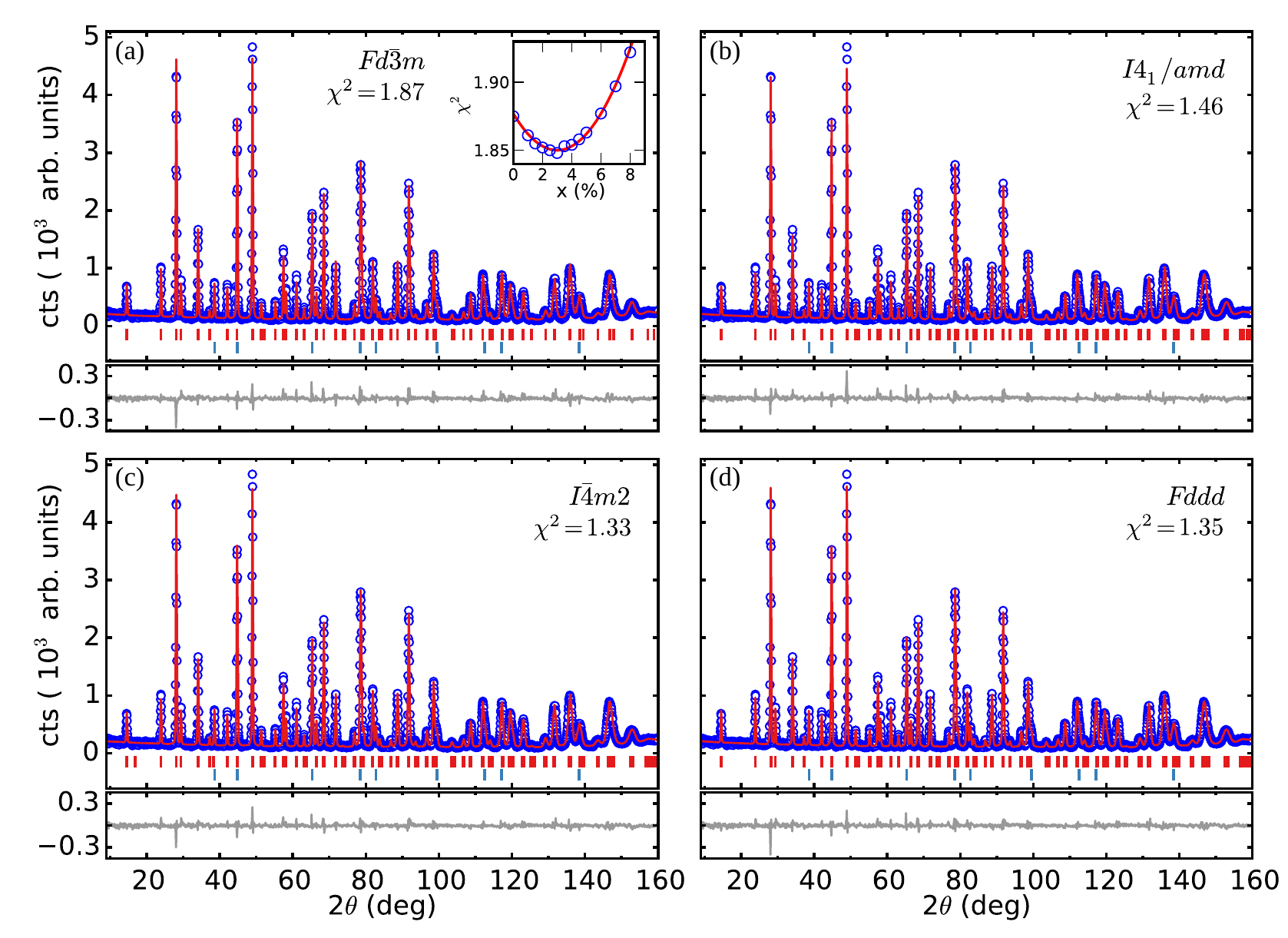}
\caption{\label{fig:Structure_refine_BT1} Structural models for \FSS{}. (a) - 
    (d) Rietveld analysis of neutron powder diffraction measured on BT-1 at 
    20~K.  Blue circles are measured data, red line is the calculated 
    diffraction pattern and red vertical bars indicate the positions of nuclear 
    reflections in the respective space group.  Contributions from the aluminum 
    sample environment and can are indicated by blue vertical bars.  The 
    difference between calculated and measured data is shown in the lower panel 
    for (a) through (d). Inset in (a) shows the goodness of fit parameter 
    ($\chi^2$) versus the Fe-Sc anti-site inversion in the cubic model.}
\end{figure*}

Neutron powder diffraction measurements collected on BT-1 at 20~K are shown in 
figure~\ref{fig:Structure_refine_BT1}; the four panels display identical 
diffraction data sets but with Rietveld refinements corresponding to different 
structural models for FSS{}. Details of each refinement are listed in 
Table~\ref{tab:struct_compare}. Note that Fe:Sc site mixing has not been 
included in these refinements, but we have checked that the conclusions here 
are not affected by this. 

\begin{table}[h!]
\caption{Candidate crystal structures for \FSS{} at 20~K.}
\begin{tabular}{ccccc}
\hline
\hline
Symmtery & cubic & tetragonal & tetragonal & orthorhombic  \\
Space group & $Fd\bar{3}m$ & $I4_1/amd$ & $I\bar{4}m2$& $Fddd$ \\
a (\AA) & 10.507(1) & 7.437(1) & 7.434(1) &10.514(1)\\
b (\AA) & / & / & / &10.514(1)  \\
c (\AA) & / & 10.495(1) & 10.493(1) &10.491(1)  \\
c/a$_{c}$  & 1 & 0.998 & 0.998 &0.998  \\
$\chi^2$  & 1.87 & 1.46 & 1.33 &1.35  \\
R$_{wp} (\%)$ & 7.92 & 6.99 & 6.68 &6.71  \\
R$_{Bragg} (\%) $ & 6.73 & 5.57 & 5.25 &5.13  \\
\hline
\hline
\end{tabular}
\label{tab:struct_compare}
\end{table}

The tetragonal lattice parameter ${\rm a}_t$ is simply related to the cubic 
cell by ${\rm a}_{t} = {\rm a}_{c}/\sqrt{2}$. Refinements for each structural 
model show that the lattice remains metrically cubic with only extremely small 
tetragonal distortions, of at most $\left({\rm a}_{c} - {\rm c}\right)/{\rm 
    a}_{c}\!=\!0.2\%$. While there are no obvious qualitative differences 
between the refinements for each structural model, close quantitative 
comparison reveals a significant improvement in the quality of fit for the 
tetragonal unit cell, and space group $I\bar{4}m2$.

\subsection{Candidate Magnetic Structures}
\label{sec:mag_models}
Irreducible representations and their basis vectors for each crystal symmetry 
and magnetic ordering wavevector in FSS{} were calculated using the SARA{\it 
    h}arah program \cite{Wills:Sarah}. Here we outline each of the candidate 
magnetic structures consistent with the neutron diffraction data in the parent 
cubic \Fdm{} unit cell and proposed tetragonal $I\bar{4}m2$ cell.
\begin{figure*}[ht!]
    \includegraphics[]{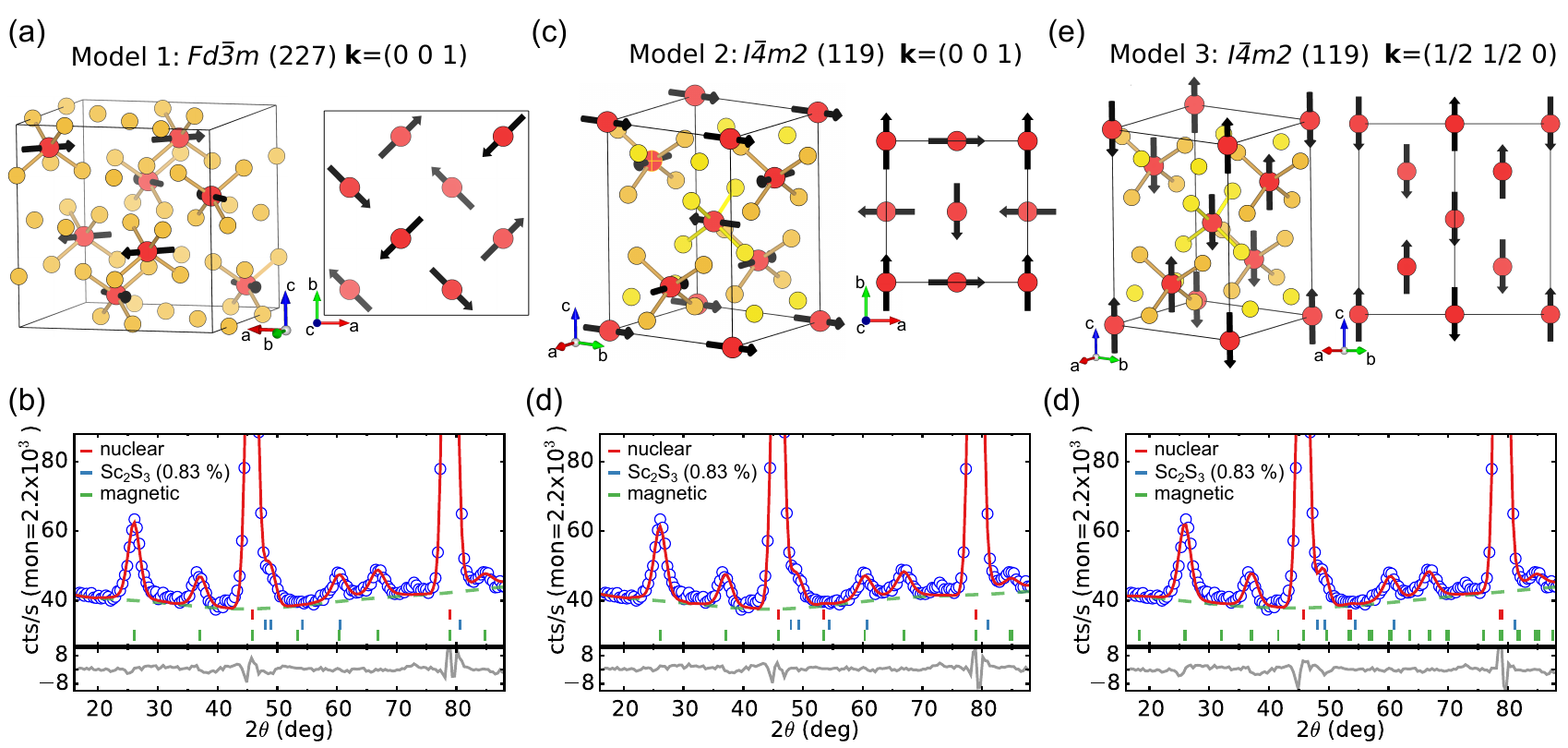}
\caption{\label{fig:altMagModels} Candidate magnetic structures in \FSS{}.  
    (a)-(b) Magnetic structure and corresponding Rietveld refinement for the 
    $Fd\bar{3}m$ structural space group and magnetic peaks indexed with a 
    $\mathbf{k} =  (0 0 1)$ propagation vector. (c)-(d) Magnetic structure and 
    corresponding Rietveld refinement for the $I\bar{4}m2$ structural space 
    group and magnetic peaks indexed with a $\mathbf{k} = (0,~0,~1)$ 
    propagation vector. The resulting magnetic structure is identical to that 
    in (a) when transformed to the cubic cell. (e)-(f) Alternative magnetic 
    structure and corresponding Rietveld refinement for the $I\bar{4}m2$ 
    structural space group and magnetic peaks indexed with a $\mathbf{k} =  
    (1/2,~1/2,~0)$ propagation vector }
\end{figure*}

\begin{figure*}[ht!]
\includegraphics[]{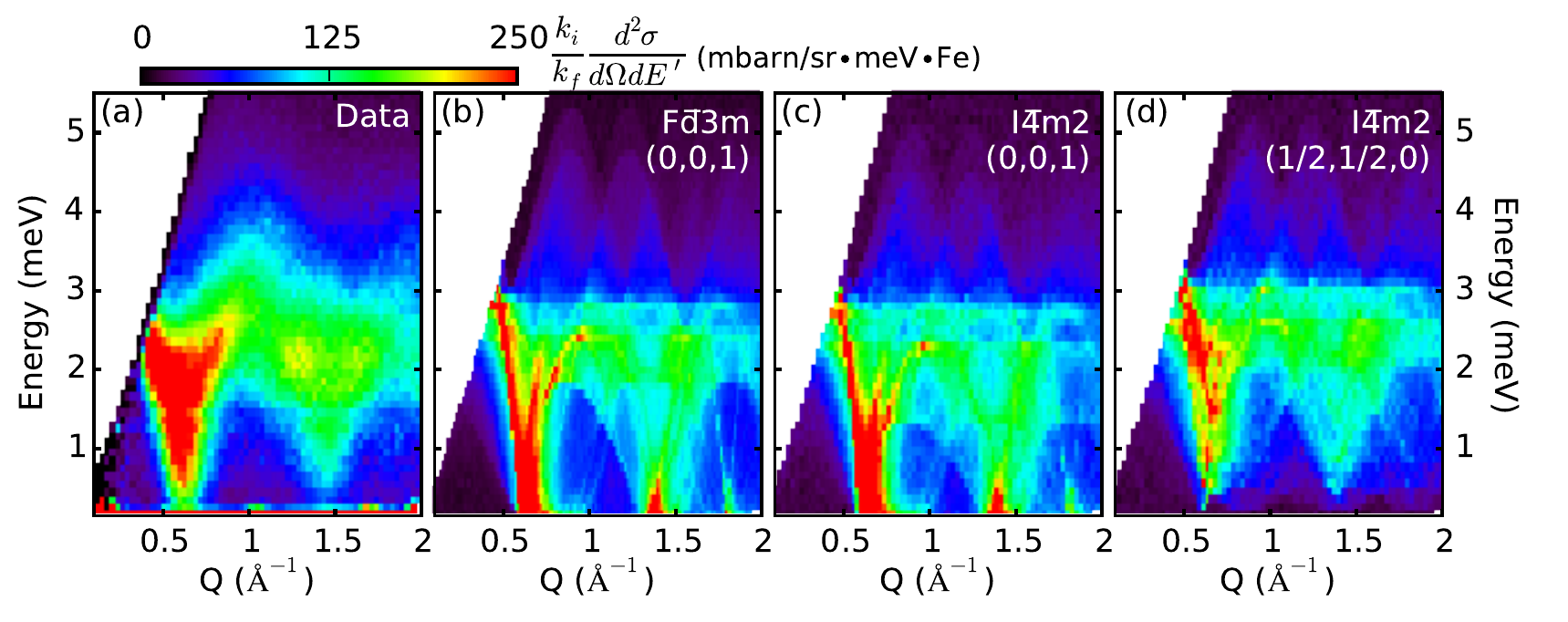}
\caption{\label{fig:SpinWModels} Spin wave excitations in candidate magnetic 
structures for \FSS{}. (a) Measured powder averaged inelastic neutron 
scattering spectra at 1.6~K and ambient pressure. (b)-(d) Calculated powder 
averaged neutron scattering intensity from spin wave excitations for
respective magnetic structures in \FSS{}. Heisenberg exchange interactions 
for each model are listed in table \ref{tab:SpinW_params}.}
\end{figure*}

In the cubic spinel structure all magnetic peaks are indexed with a propagation 
vector of ${\bf k}=( 0,~ 0,~ 1)$. The decomposition of the magnetic 
representation for the Fe site $(0.125,~0.125,~0.125)$ is 
$\Gamma_{Mag}=\Gamma_{1}^{2}+\Gamma_{2}^{2}+2\Gamma_{4}^{2}$, the irreducible 
representations and associated basis vectors are detailed in 
Table~\ref{table:227_irreps}.  Satisfactory magnetic refinements in the \Fdm{} 
structural space group are only possible including one equally weighted basis 
vector from each of the two irreducible representations $\Gamma_{1}$ and 
$\Gamma_{2}$: either $\bfpsi_{1}$ and $\bfpsi_{3}$ or $\bfpsi_{2}$ and 
$\bfpsi_{4}$.  The two structures are simply related by a $\pi/2$ rotation of 
all spins and are indistinguishable in powder diffraction data. The resulting 
magnetic structure and refinements  are shown in Fig.~\ref{fig:altMagModels} 
(a).
\begin{table}[h!]
\caption{Basis vectors for the space group \Fdm{} with ${\bf k}=( 0,~ 0,~ 1)$. 
}
\begin{tabular}{cccccc}
\hline
\hline
  Irrep.  &  Basis Vector  &  Atom & $m_{\|a}$ & $m_{\|b}$ & $m_{\|c}$ \\
\hline
$\Gamma_{1}$ & $\bfpsi_{1}$ &      1 &      0 &      4 &      0   \\
             &              &      2 &     -4 &      0 &      0   \\
             & $\bfpsi_{2}$ &      1 &      4 &      0 &      0   \\
             &              &      2 &      0 &     -4 &      0   \\
\hline
$\Gamma_{2}$ & $\bfpsi_{3}$ &      1 &      4 &      0 &      0   \\
             &              &      2 &      0 &      4 &      0   \\
             & $\bfpsi_{4}$ &      1 &      0 &     -4 &      0   \\
             &              &      2 &     -4 &      0 &      0   \\
\hline
$\Gamma_{4}$ & $\bfpsi_{5}$ &      1 &      0 &      0 &      8   \\
             &              &      2 &      0 &      0 &      0   \\
             & $\bfpsi_{6}$ &      1 &      0 &      0 &      0   \\
             &              &      2 &      0 &      0 &     -8   \\
\hline
\hline
\end{tabular}
\label{table:227_irreps}
\end{table}

For the tetragonal $I\bar{4}m2$ cell the magnetic Bragg peaks may all be 
indexed with a magnetic propagation vector of  either ${\bf k_1}=( 0,~ 0,~ 1)$.  
or ${\bf k_2}=(1/2,~1/2,~0)$. There are two Fe sites in the primitive basis, 
for the Fe$_1$ at $( 0,~ 0,~ 0)$ the decomposition of the magnetic 
representation is $\Gamma_{Mag}=\Gamma_{2}^{1} + \Gamma_{5}^{2}$ for ${\bf 
    k_1}$ and $\Gamma_{Mag}=\Gamma_{2}^{1}+\Gamma_{3}^{1}+\Gamma_{4}^{1}$ for 
${\bf k_2}$. On the Fe$_2$ site at $(0,0.5,0.25)$, the decomposition of the 
magnetic representation is $\Gamma_{Mag}=\Gamma_{4}^{1}+\Gamma_{5}^{2}$ for 
${\bf k_1}$ and $\Gamma_{Mag}=\Gamma_{1}^{1}+\Gamma_{2}^{1}+\Gamma_{4}^{1}$ for 
${\bf k_1}$. Irreducible representations and associated basis vectors for the 
propagation vectors ${\bf k_1}$ and ${\bf k_2}$ are detailed in tables 
\ref{table:119_irreps1a}, \ref{table:119_irreps1b} and 
\ref{table:119_irreps2a}, \ref{table:119_irreps2b} respectively.

The refined magnetic structure for a ${\bf k_1}=( 0,~ 0,~ 1)$ propagation 
vector is  shown in figure~\ref{fig:altMagModels} (c) and is identical to that 
in figure~\ref{fig:altMagModels} (a) when transformed into the cubic cell. This 
structure is described with the irreducible representation $\Gamma_{5}$ on both 
Fe sites and equally weighted basis vectors. 

\begin{table}[h!]
\caption{Basis vectors for the space group $I\bar{4}m2$ with ${\bf k} =( 0,~ 
    0,~ 1)$ for for the $Fe$ site $(0,~ 0,~ 0)$}
\begin{tabular}{cccccc}
\hline
\hline
  Irrep.  &  Basis Vector  &  Atom & $m_{\|a}$ & $m_{\|b}$ & $m_{\|c}$ \\
\hline
$\Gamma_{2}$ & $\bfpsi_{1}$ &      1 &      0 &      0 &      8 \\
\hline
$\Gamma_{5}$ & $\bfpsi_{2}$ &      1 &      0 &      4 &      0 \\
             & $\bfpsi_{3}$ &      1 &      4 &      0 &      0 \\
\hline
\hline
\end{tabular}
\label{table:119_irreps1a}
\end{table}
\begin{table}[h!]
\caption{Basis vectors for the space group $I\bar{4}m2$ with ${\bf k}=( 0,~ 0,~ 
    1)$ for the $Fe$ site $( 0,~ 0.5,~ 0.25)$.}
\begin{tabular}{cccccc}
\hline
\hline
  Irrep.  &  Basis Vector  &  Atom &$m_{\|a}$ & $m_{\|b}$ & $m_{\|c}$\\
\hline
$\Gamma_{4}$ & $\bfpsi_{1}$ &      1 &      0 &      0 &      8  \\
\hline
$\Gamma_{5}$ & $\bfpsi_{2}$ &      1 &      0 &      4 &      0  \\
             & $\bfpsi_{3}$ &      1 &     -4 &      0 &      0  \\
\hline
\hline
\end{tabular}
\label{table:119_irreps1b}
\end{table}
When the magnetic peaks are indexed with ${\bf k_1}=( 1/2,~1/2,~0)$ many 
equally good solutions are permitted by the data. In 
figure~\ref{fig:altMagModels} (e) and (d) we show the magnetic structure and 
refinement for  $\Gamma_4$ on Fe$_1$ and $\Gamma_2$ on Fe$_2$.  

The collinear magnetic structure shown in figure~\ref{fig:MagStructure} (b) and 
(d) of the main text includes irreducible representations $\Gamma_3$ on Fe$_1$ 
and $\Gamma_1$ on Fe$_2$.
\begin{table}[h!]
\caption{Basis vectors for the space group I$\bar{4}m2$ with ${\bf k} =( 1/2,~ 
    1/2,~ 0)$ for the $Fe$ site $( 0,~ 0,~ 0)$.}
\begin{tabular}{cccccc}
\hline
\hline
  Irrep.  &  Basis Vector  &  Atom &$m_{\|a}$ & $m_{\|b}$ & $m_{\|c}$ \\
\hline
$\Gamma_{2}$ & $\bfpsi_{1}$ &      1 &      2 &     -2 &      0  \\
$\Gamma_{3}$ & $\bfpsi_{2}$ &      1 &      2 &      2 &      0  \\
$\Gamma_{4}$ & $\bfpsi_{3}$ &      1 &      0 &      0 &      4  \\
\hline
\hline
\end{tabular}
\label{table:119_irreps2a}
\end{table}
\begin{table}[h!]
\caption{Basis vectors for the space group  $I\bar{4}m2$ with ${\bf k}=( 1/2,~ 
    1/2,~ 0)$ for the $Fe$ site $( 0,~ 0.5,~ 0.25)$ .}
\begin{tabular}{cccccc}
    \hline
    \hline
  Irrep.  &  Basis Vector  &  Atom &$m_{\|a}$ & $m_{\|b}$ & $m_{\|c}$ \\
\hline
$\Gamma_{1}$ & $\bfpsi_{1}$ &      1 &      2 &      2 &      0  \\
$\Gamma_{2}$ & $\bfpsi_{2}$ &      1 &      0 &      0 &      4  \\
$\Gamma_{4}$ & $\bfpsi_{3}$ &      1 &      2 &     -2 &      0  \\
\hline
\hline
\end{tabular}
\label{table:119_irreps2b}
\end{table}\\

\subsection{Magnetic Excitations and Spin Wave Models}
\label{sec:spinwaves}
The powder averaged magnetic excitation spectra was computed for each magnetic 
candidate structure in \FSS{} within linear spin wave theory as implemented in 
the \texttt{SpinW} program \cite{Toth:spinw}. For each model, the inelastic 
magnetic neutron scattering intensity from spin waves was calculated using a 
simple model Hamiltonian, including only Heisenberg type exchange terms 
labelled in Fig.~\ref{fig:MagStructure} (b) of the main text as well as single 
ion anisotropy terms on each Fe$^{2+}$. For the cubic structure a symmetry 
disallowed easy plane anisotropy term was included to stabilize the magnetic 
structure; while the tetragonal structures allow for two distinct single ion 
anisotropy terms: one per crystallographic Fe site.  These were included as 
easy plan anisotropies for models 2 and 4, and an easy axis anisotropy for 
model 3. Models 1 through 3 are shown in figure~\ref{fig:altMagModels} and 
model 4 is that presented in the main text.
\begin{table}[h!]
\caption{Exchange interactions used for model spin wave calculations in \FSS{}.  
    Bonds are labeled according to Fig. \ref{fig:MagStructure} (b) in the main 
    text.}
\begin{tabular}{c|ccc}
\hline
\hline
 \multicolumn{1}{c}{}& Model 1 & Model 2  & Model 3  \\
 \hline
$J_1$ (meV)      & 0.10(9) & 0.10(9) & -0.04(4) \\
$J_2$ (meV)      & 0.27(5) & 0.27(5) & 0.32(6) \\
$J_2^{\prime}$ (meV)  & -- & 0.27(5) & 0.27(6) \\
$J_3$ (meV)        & 0.00(1) & 0.00(2) &  0.01(1) \\
$J_3^{\prime}$ (meV)  & -- & 0.00(2) & 0.01(1)\\
$J_4$ (meV)        & 0.00(1) & -0.01(2) & -0.04(3)\\
$J_4^{\prime}$ (meV) & -- & -0.01(2) &-0.04(3)\\
$D$ (meV)          & 0.1(1) & 0.15(6) & 0.15(6)\\
$D^{\prime}$ (meV) & -- & 0.05(4) &0.05(4)\\
\hline
\hline
\end{tabular}
\label{tab:SpinW_params}
\end{table}
Guided by the relative strength of spin correlators and magnetic bond energies 
extracted from the static structure factor and first moment sum rule analysis, 
the magnetic exchange interactions where chosen to provide the best qualitative 
match of the data with the simplest combination of exchange terms. Exchange 
constants used for each respective spin wave calculation are detailed in 
Table~\ref{tab:SpinW_params}.

The ground state energy of each magnetic structure are as follows:
\begin{align*}
Model\,\,1:& &E_1 &= 0J_1 -8J_2-4J_4, \\
Model\,\,2:& &E_2 &= 2\!\times\!\left(0J_1+4J_2-8J_2^{\prime}-2J_4\right),\\
Model\,\,3:& &E_3 &= 2\!\times\!\left(0J_1-4J_2+0J_2^{\prime}-2J_4\right).
\end{align*}
Ground state energies of models 3 and 4 are identical. For these models 
Heisenberg NN ($J_1$) and out-of-plane NNN ($J_2^{\prime}$) exchange terms 
cancel at the mean field level and the systems is effectively a 
\emph{two-dimensional} Heisenberg square lattice.  For Models 1 and 2 the 
nearest neighbor spins are orthogonal and Heisenberg terms have no effect.

%
\end{document}